\documentclass[english,12pt,nofootinbib,preprintnumbers,showpacs,eqsecnum,prd]{revtex4}

\usepackage{graphicx}
\usepackage{amssymb}
\usepackage{babel}
\usepackage{epsfig}

\newcommand \beq{\begin{eqnarray}}
\newcommand \eeq{\end{eqnarray}}
\def\simle{\mathrel{\rlap{\raise 0.511ex \hbox{$<$}}{\lower 0.511ex 
\hbox{$\sim$}}}}
\def\simge{\mathrel{ \rlap{\raise 0.511ex 
\hbox{$>$}}{\lower 0.511ex \hbox{$\sim$}}}}

\begin{document}

\preprint{ECT{*}-06-15}

\title{Perturbation theory and non-perturbative renormalization flow\\ in scalar field theory at finite temperature}

\author{Jean-Paul Blaizot}
\affiliation{ECT{*}, Villa Tambosi, Strada delle Tabarelle 286, \\
I-38050 Villazzano Trento, Italy}

\author{Andreas Ipp}
\affiliation{ECT{*}, Villa Tambosi, Strada delle Tabarelle 286, \\
I-38050 Villazzano Trento, Italy}

\author{Ram\'on M\'endez-Galain}
\affiliation{Instituto de F\'\i sica, Facultad de Ingenier\'\i a,\\
J.H.y Reissig 565, 11000 Montevideo, Uruguay}

\author{Nicol\'as Wschebor}
\affiliation{Instituto de F\'\i sica, Facultad de Ingenier\'\i a,\\
J.H.y Reissig 565, 11000 Montevideo, Uruguay}

\date{28th November 2006}

\begin{abstract}
We use the non-perturbative  renormalization group to clarify some features
of perturbation theory in thermal field theory. For the specific
case of the scalar field theory with $O(N)$ symmetry, we solve the
flow equations within the local potential approximation. This
approximation reproduces the perturbative results for the screening 
mass and the pressure up to
order $g^3$, and starts to differ at order $g^4$. The method allows a
smooth extrapolation to the regime where the coupling is not small, 
very similar to that obtained from a  simple
self-consistent approximation.
\end{abstract}

\maketitle

\section{Introduction}

Perturbative calculations at high temperature suffer from bad
convergent behavior. Even the thermodynamical quantities, such as
the pressure or the entropy density, which are dominated by short
wavelength degrees of freedom -- for which one could expect
perturbation theory to be reasonably good -- are not well described
by a strict expansion in powers of the coupling constant. This
problem has been particularly studied in quantum chromodynamics, in
the context of the quark-gluon plasma  (for a recent review, see
\cite{Blaizot:2003tw}). In this case, the perturbative expansion of
the thermodynamic potential is known up to order $g^{6}\log g$,
where $g$ is the gauge coupling
\cite{Arnold:1995eb,Zhai:1995ac,Braaten:1996jr,Kajantie:2002wa} (the
$g^{6}$ term requires non-perturbative methods
\cite{DiRenzo:2006nh}). However, the problem is not specific to QCD
at high temperature: Similar poor convergence behavior appears also
in the simpler scalar field theory \cite{Parwani:1995zz}, and has also
been observed in the case of large-$N$ $\phi^{4}$ theory
\cite{Drummond:1997cw}.

Reorganizations of the perturbative expansion, based on various
arguments, have been proposed in order to extend the usefulness of
weak coupling calculations to regimes of not too small coupling.
These involve ``screened perturbation theory'' \cite{Karsch:1997gj},
``hard thermal loop perturbation theory''
\cite{Andersen:2000yj,Andersen:1999fw,Andersen:2002ey,Andersen:2003zk,Andersen:2004fp},
and the approach put forward in
Refs.~\cite{Blaizot:1999ip,Blaizot:1999ap,Blaizot:2000fc}, based on 
an expansion of the thermodynamical potential in terms of dressed 
propagators which involve two-particle irreducible (2PI) diagrams 
\cite{Luttinger:1960ua,Baym:1962sx,Cornwall:1974vz}.
The approach of \cite{Blaizot:1999ip,Blaizot:1999ap,Blaizot:2000fc} 
exploits a nonperturbative expression for the entropy
density that can be obtained from a $\Phi$-derivable two-loop
approximation \cite{Vanderheyden:1998ph}. Here the emphasis is on a
physical picture involving quasiparticles whose residual
interactions are assumed to be weak after the bulk of the
interaction effects have been incorporated in the spectral
  properties of these quasiparticles. It has been shown in particular 
that the entropy
density of the quark gluon plasma can be well understood within such
a scheme down to temperatures $T\gtrsim3T_{c}$ where the coupling
can be as
large as $g\approx2$ \cite{
Blaizot:2000fc}. As an alternative to resummations,
effective field theories have also been used \cite{Braaten:1996jr}.
Among those, dimensional reduction, which emphasizes the role of the
zero Matsubara frequency, stands out as a very powerful one 
\cite{Kajantie:2000iz,Braaten:1995cm}. An interesting feature of 
the perturbative
calculations when organized through the dimensionally reduced
effective theory is that the large scale dependence
of strict perturbation theory is considerably reduced when the
effective parameters are not subsequently expanded out
\cite{Kajantie:2002wa,Blaizot:2003iq}. When this is done,
the predictions of dimensional reduction become similar to those of
the 2PI resummation.

In fact, the origin of the difficulties of thermal perturbation
theory is now well understood. What complicates the situation is the
fact that the coupling constant alone does not control the magnitude
of corrections to the free theory: Thermal fluctuations of various
wavelengths play also an essential role. In the weak coupling
regime, the thermal system is characterized by a hierarchy of scales: 
Most particles have momenta of
the order of the temperature. At high temperature, these ``hard''
degrees of freedom dominate the thermodynamics and their
interactions are accurately described by ordinary perturbation
theory. However, perturbative corrections to the thermodynamical
functions involve also ``soft'' degrees of freedom, whose momenta
are typically of order $gT$. These soft modes are non-perturbatively
related by their coupling to the hard modes. Such corrections
are easily handled by resummations or effective theories.
Furthermore, the soft modes also interact among themselves. Although
perturbative, these interactions generate corrections which take the
form of an expansion in powers of $g$ rather than of $g^2$ as is the
case for ordinary perturbation theory: as a result, perturbation
theory becomes less accurate.
In QCD, there exists a further scale, $g^2T$ for which perturbation
theory stops to make sense: at this scale, the self interactions of
the modes become comparable to their kinetic energies, invalidating any
expansion around free particle motion. The problems associated with
this ``ultra-soft'' scale are specific to non-abelian gauge theories;
they will not be discussed in this paper.

There are indications, already alluded to above, that the structure 
identified in weak
coupling, which is based on the hierarchy of scales that we have
just discussed,  seem to survive, after appropriate resummation, or
appropriate use of effective theories, even in regimes of strong
coupling where the arguments used to identify the structure become 
unjustified. That is, the resummations that have been motivated by 
analyzing the weak coupling regime, provide a smooth extrapolation 
into the regime of strong coupling where a strict expansion in powers 
of the coupling does not make sense.
  The purpose of this paper is to
shed light on this issue by using the non-perturbative renormalization
group (NPRG)
\cite{Wetterich:1992yh,Ellwanger:1993kk,Tetradis:1993ts,Morris:1993qb,Morris:1994ie}.

There is some analogy between the effective field theory approach
and the NPRG: in effective field theory
one integrates out degrees of freedom above some cut-off; in the
renormalization group this integration is done smoothly. In a sense,
the renormalization group builds up a continuous tower of effective
theories that lie infinitely close to each other and are labeled by
a momentum cut-off scale $\kappa$. These effective theories are
related by a renormalization group flow equation. The picture
remains essentially the same for any value of the coupling.

The NPRG has been applied, in various
incarnations (it is also called exact, or functional),  to a variety 
of problems in condensed matter
\cite{Delamotte:2003dw,Delamotte:2004zg}, 
in particle
\cite{Ellwanger:1993mw,Ellwanger:1994wy,Ellwanger:1996wy,Pawlowski:2003hq,Fischer:2004uk,%
Arnone:2005fb,Morris:2006in}
and nuclear physics (for reviews see \cite{Bagnuls:2000ae,Berges:2000ew,Pawlowski:2005xe}).
It has been applied to problems at finite temperature 
\cite{Tetradis:1993ts,Reuter:1993rm,Andersen:1999dy,D'Attanasio:1996zt,Liao:1995gt,Braun:2003ii}, 
and
the general behaviors that we shall report in this paper have been
known already for some time. However, the main focus of previous
studies has been the description of the phase transitions, rather
than the specific problem that we want to address here.

  The outline of this paper is as follows. In Sec.~\ref{sec:pert_theory}, we
present a brief review of available results concerning perturbation
theory for a scalar field with $O(N)$ symmetry. We also discuss a
simple self-consistent $\Phi$-derivable approximation that becomes
exact in the large $N$ limit. Section \ref{sec:RGandLPA} gives a
brief introduction to the non-perturbative renormalization group and the
local potential approximation. Specific features of finite
temperature calculations are recalled.  In particular we introduce a
regulator that we found particularly convenient for such
calculations.  In Sec.~\ref{sec:numerical} we integrate the flow
equations numerically and discuss the  results obtained.
  Finally, in Section~\ref{sec:perturbative},  we perform a perturbative analysis of the
flow equations and comment the results. Note that a preliminary
account of this work was presented in Refs.~\cite{Blaizot:2006xx, Blaizot:2006xy}.

\section{Perturbation theory and the need for 
resummation\label{sec:pert_theory}}

In this section we briefly review existing results of perturbation
theory for the thermodynamics of the scalar field. We also recall
how a simple self-consistent approximation  based on the lowest oder 
2PI diagram, can be used to include
non perturbative effects that allow  for a smooth extrapolation to
strong coupling. This will be useful later in our discussion, as it
turns out that the results of this self-consistent approximation are
close to those of the non-perturbative RG within the local potential
approximation.

\subsection{Brief review of results from perturbation theory}
\label{sec:rev-pert}
We consider a scalar $\phi^{4}$ theory in $d$ dimensions, defined by the $O(N)$
symmetric Lagrangian
\begin{equation}
\mathcal{L}(x)=\frac{1}{2}\left[\partial\varphi(x)\right]^{2}-\frac{1}{2}m^{2}\varphi^{2}(x)-g^{2}\left[\varphi^{2}(x)\right]^{2}
\label{eq:classicalL},
\end{equation}
where the field $\varphi(x)$ has $N$ real components $\varphi_{i}(x)$,
with $i=1,...,N$.
The temperature $T$ ($\beta = 1/T$) enters through 
the action
\begin{equation}
S=\int_{0}^{\beta}d\tau\int d^{d-1}x \, \mathcal{L}(x) ,
\label{eq:classicalS}
\end{equation}
and the periodicity of the fields in imaginary time, $\varphi(\beta) 
= \varphi(0)$. 
We shall fix the bare parameters of the action at $T=0$, in order
to have a vanishing renormalized mass and a given value of the 
renormalized coupling constant (the last point is discussed in Sec.~\ref{sec:towardssc}).
Then we turn on the temperature at a given fixed value and we study the finite temperature predictions 
of the model for various values of the zero-temperature coupling.

There are two physical quantities that we shall discuss in this 
paper: the screening mass\footnote{Note that the screening mass, 
defined as the pole of the static propagator
(for complex wave numbers), differs from the quasiparticle mass,
defined as the pole of propagator at vanishing three-momentum. Up to 
order $g^3$, the two quantities coincide in scalar theory
but start to differ at order $g^4$. Furthermore, within the 
approximations that we shall consider in this paper, where the 
momentum dependence of the self-energy is neglected, these two masses 
are  the same, and coincide also with the second derivative of the 
effective potential with respect to the field.}, and the pressure.
\begin{figure}
\begin{center}\includegraphics[%
   scale=0.7]{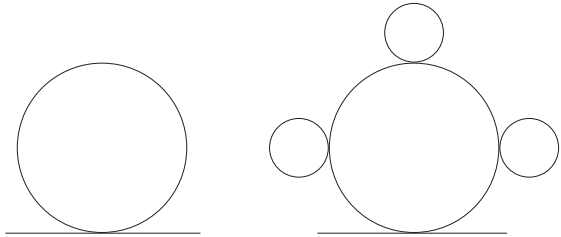}\end{center}
\caption{Diagrams that contribute to the screening mass $m_D^{2}$
  at order  $g^{2}$ (left) and at order  $g^{3}$ (right).\label{cap:Tadpole}}
\end{figure}
%
  Up to  order $g^4$,
the screening mass reads \cite{Braaten:1995cm}:
\beq \label{mpert}
\!\!m_D^2=g^2T^2\Big\{1- 3 \frac{g}{\pi} -\frac{9}{2} \left[
\ln{\bar\mu\over2\pi T}- {4\over 3}\ln {g\over \pi}-2.415 \right]
\!\left(\frac{g}{\pi}\right)^2+{\mathcal O}(g^3)\Big\} ,
\eeq
where $\bar\mu^2 = 4\pi e^{-\gamma} \mu^2$ with $\mu^2$ the scale 
introduced in dimensional regularization,
$\gamma$ the Euler constant, and here $g^2 = g^2(\bar\mu)$ is the $\overline{\rm 
MS}$ coupling constant
at the renormalization scale $\bar\mu$. 
Similarly, the 
weak--coupling expansion of the pressure has been computed (in
the massless case) to order $g^5$, and reads
\cite{Arnold:1994ps,Parwani:1995zz,Braaten:1995cm}:
\beq\label{Pphi}\nonumber P &=& P_0\Bigg[ 1-{15\over
8}\left(\frac{g}{\pi}\right)^2 +{15\over
2}\left(\frac{g}{\pi}\right)^3 +{135\over 16}
\left(\log{\bar\mu\over2\pi
T}+0.4046\right)\left(\frac{g}{\pi}\right)^4
\\
&{}&\quad -{405 \over 8}\left(\log{\bar\mu\over2\pi
T}-{4\over3}\log\frac{g}{\pi} -0.9908
\right)\left(\frac{g}{\pi}\right)^5 +{\mathcal O}(g^6 \log
g)\Bigg]\;, \eeq where $P_0 = (\pi^2/90)T^4$ is the pressure of an
ideal gas of free massless bosons. 

  The lack of convergence of
the weak-coupling expansion of both the screening mass and the 
pressure, manifests itself by the fact that, unless the coupling is 
very small, $g\simle .1$,  the successive corrections do not decrease 
in magnitude, and the dependence of the renormalization scale gets 
larger and larger (see e.g.~\cite{Blaizot:2003tw}). This is in contrast to what one would expect in  a 
well behaved perturbative expansion where the explicit scale 
dependence  cancels against that of the running coupling, to within 
terms of higher order than those retained in the calculation. Such 
cancellations do indeed occur in Eqs.~(\ref{mpert}) and (\ref{Pphi}). 
To see that,  it is enough to use the one-loop $\beta$ 
function:
\begin{equation}
\mu\frac{\partial g^{2}}{\partial\mu}\,=\,\frac{9}{2\pi^{2}} g^{4} + O(g^6).
\label{eq:betafunction}
\label{renorm0}
\end{equation}
However, since the successive orders do not get 
smaller and smaller, the cancellation of the scale dependence remains 
only formal.

\begin{figure}
\begin{center}\includegraphics[%
   scale=0.7]{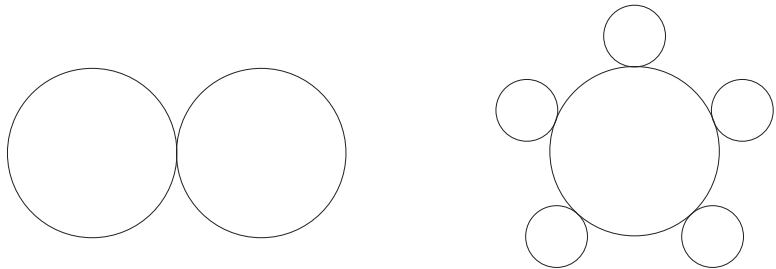}\end{center}
\caption{Diagrams contributing to the pressure at order $g^{2}$ (left diagram)
and $g^{3}$ (right diagram).\label{cap:pressurediagrams}}
\end{figure}

It is instructive to recall the origin of the first 
two terms of the expansions of the screening mass and the pressure
(which can be obtained from the diagrams depicted in Figs.~\ref{cap:Tadpole} and \ref{cap:pressurediagrams}), 
because they illustrate well some aspects of the physics that we are 
discussing. 
We shall work out only the calculation of the mass; that 
of the pressure is similar.

The first two terms in Eq.~(\ref{mpert}) are obtained from the
Feynman diagrams displayed in Fig.~\ref{cap:Tadpole} which we
calculate as (using an ultraviolet (UV) cut-off rather than dimensional
regularization for the vacuum part)
\begin{eqnarray}
m_D^{2} & = &
12 g^2 T \sum_{\omega_n} \int\frac{d^3 q}{(2\pi)^3} 
\frac{1}{\omega_n^2 + q^2 + m^2 } + \delta m^2
\nonumber \\
& = &
12g^{2}\int_{0}^{\Lambda}\frac{q^{2}dq}{2\pi^{2}}\frac{1+2n(\omega_q)}{2\omega_q}+\delta 
m^{2}
\,,\label{eq:pert01}
\end{eqnarray}
with $\omega_q^2 = q^2 +m^2$. In the second line of
(\ref{eq:pert01}) the sum over Matsubara frequencies $\omega_n = 2
\pi n T$ has been converted into an integral over the distribution
function $n(\omega)=(e^{\omega/T}-1)^{-1}$ (see Appendix A), and a
UV regulator $\Lambda$ introduced. The counterterm
$\delta m^{2}$, of order $g^{2}$, cancels the ultraviolet divergence
(as $\Lambda \rightarrow \infty$) of the vacuum integral and its 
finite part is chosen so that the renormalized mass vanishes in the 
vacuum. 

The order
$g^2$ contribution is a genuine perturbative correction, dominated by 
the hard degrees of freedom. It can be estimated by  neglecting the 
mass in $\omega_q$:
\begin{eqnarray}
m_D^{2} & = &
\frac{6g^{2}}{\pi^{2}}\int_{0}^{\infty}dq\, qn(q) = g^{2}T^{2}\,.
\label{eq:pert14}
\label{eq:pert15}
\end{eqnarray}

  The $g^{3}$ term however is not, strictly speaking, a perturbative 
correction: the odd power is the result of an infinite resummation. 
What is at work here is precisely the coupling between soft modes and 
hard ones that we discussed in the introduction: when the momentum 
running in the loop of the left diagram of Fig.~\ref{cap:Tadpole} is 
soft, i.e., of order $gT$, the correction to the propagator due to 
the hard fluctuations cannot be ignored since  these hard 
fluctuations contribute a mass also of order $gT$.  Thus one needs to 
keep the thermal mass in the
propagator. Starting from (\ref{eq:pert01}) and
subtracting the $O(g^2)$ contribution that we have just calculated,
we obtain
\begin{eqnarray}
\left. m_D^{2} \right|_{g^3} & = & 
\frac{6g^{2}}{\pi^{2}}\int_{0}^{\infty}q^{2}dq\left(\frac{n(\omega_q)}{\omega_q}-\frac{n(q)}{q}\right) 
\nonumber \\
  & = & 
\frac{6g^{2}}{\pi^{2}}\int_{0}^{\infty}q^{2}dq\left(\frac{T}{\omega_q^{2}}-\frac{T}{q^{2}}\right)
  \nonumber \\
  & = & -\frac{3}{\pi}g^{3}T^{2} .
\label{eq:pertg3}
\end{eqnarray}
In the second line we have used the approximated form of the 
statistical factors
$n(\omega) \simeq T/\omega$,
appropriate since the integral is dominated by soft momenta $q \ll
T$. (One would have obtained the same second line of
(\ref{eq:pertg3}) by starting from the  expression in the first line
of (\ref{eq:pert01}) and keeping only the $\omega_n = 0$
contribution.)

The type of resummation involved to get the $g^3$ term can be turned 
into a fully self-consistent approximation which extrapolates smoothly 
to strong coupling. But before we do that, let us add a few remarks 
about the running coupling, and in particular about what is meant by 
strong coupling in the present discussion.

\subsection{Remarks about the coupling}
\label{sec:towardssc}

The running of the coupling, to order one-loop, is governed by the
$\beta$-function (\ref{eq:betafunction}). Assuming an ultraviolet 
cut-off $\Lambda$, where the value of the coupling is $g_\Lambda$ and 
integrating  Eq.~(\ref{eq:betafunction}), one gets: 
\begin{equation}
\frac{1}{g_{\mu}^{2}}=\frac{1}{g_{\Lambda}^{2}}+\frac{9}{2\pi^{2}}\log\frac{\Lambda}{\mu} 
,
\label{eq:couplingrelation}
\end{equation}
which shows that $g_\mu\to 0$ when $\mu\to 0$. The  coupling at
scale $\mu = 0$ is therefore not a suitable quantity to characterize
the theory. In the following, we therefore adopt the usual
practice of fixing the coupling at
the scale $2 \pi T$, which we shall simply denote by $g\equiv g(\mu=2\pi
T)$ (unless ambiguities may arise). 
Equation~(\ref{eq:couplingrelation}) can then be rewritten as 
\begin{equation}
\frac{1}{g_{\mu}^{2}}=\frac{1}{g^{2}}+\frac{9}{2\pi^{2}}\log\frac{2\pi 
T}{\mu} \,.
\label{eq:couplingrelationvac}
\label{eq:couplingrelation2}
\end{equation}
Now,  for any finite value of  $g$ there is a scale $\mu = 
\Lambda_{{\rm L}}$ (the ``Landau pole'')
at which the coupling diverges:
\begin{equation}
\Lambda_{{\rm L}}=2\pi T\exp\frac{2\pi^{2}}{9g^2}\,.
\label{eq:LandauPole}
\end{equation}
If $g \ll 1$ then $\Lambda_{\rm L }\gg \Lambda$ and the
physics is not affected by $\Lambda_{\rm L}$.
However, as $g$ grows, $\Lambda_{\rm L}$ decreases and eventually 
becomes of same order as $\Lambda$. In order to avoid unphysical 
behaviors, we require $\Lambda_{\rm L}>\Lambda$, which puts a 
constraint on the largest admissible values of $g$:
 \begin{equation}
g^{2}<\frac{2\pi^{2}}{9\log(\Lambda/2\pi T)} \,.
\label{eq:couplingbound}
\end{equation}
 It is easy to verify that this bound corresponds to 
an infinite value of $g_\Lambda$. To increase the maximum value of 
$g$, one could increase the temperature; however if $2\pi T$ becomes 
too close to $\Lambda$, the results become sensitive to the value of 
the cut-off (see the section on numerical results for further 
discussion of this point). Note that 
  this estimate of the upper 
bound for $g$
is modified by non-perturbative effects, as we shall see 
in Sec.~\ref{sec:numerical}.

\begin{figure}
\begin{center}\includegraphics[%
   scale=0.8]{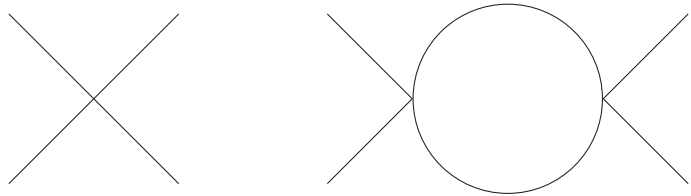}\end{center}
\caption{Diagrams contributing to the four-point function at order $g^{2}$
(left diagram) and $g^{4}$ (right diagram).\label{cap:couplingdiagrams}}
\end{figure}

The coupling constant may be viewed as the value of the scattering
amplitude $A$ for vanishing external momenta. While, as we have just
seen, after renormalization this scattering amplitude vanishes, this
is not so at finite temperature. Up to order $g^{4}$, $A$ can be 
calculated from the diagrams in
Fig.~\ref{cap:couplingdiagrams}. We obtain
\begin{equation}
A = 
g^{2}+36g^{4}T\sum_{\omega_{n}}\int\frac{d^{3}q}{(2\pi)^{3}}\left(\frac{1}{\omega_{n}^{2}+q^{2}+m_D^{2}}\right)^{2} 
.
\label{eq:scatteringamplitude1}
\end{equation}
We can  calculate the sum-integral using the formulae given in 
Appendix A or,  more directly, noticing that the thermal
contribution is infrared dominated, by keeping only the contribution 
of  $\omega_n = 0$ in
(\ref{eq:scatteringamplitude1}). We obtain
\begin{eqnarray}
A & \approx & 
g^{2}-\frac{18g^{4}}{\pi^{2}}\int_{0}^{\infty}q^{^{2}}dq 
\frac{T}{\omega_{q}^{4}} =g^{2}-\frac{9g^{3}}{2\pi} ,
  \end{eqnarray}
  where $g^2 = g^2(2\pi T)$. Thus the scattering amplitude contains a 
$g^3$ contribution, whose origin is the same as that of the $g^3$ 
contributions in the mass or the pressure.

\subsection{Self-consistent 2PI resummation}

 As we have seen earlier, resummations are required in order to 
deal with the IR aspects of thermal perturbation theory. We shall 
briefly recall here the equations that are obtained in the simple 
$\Phi$--derivable approximation where the single 2PI diagram that is 
considered is the left diagram in Fig.~\ref{cap:pressurediagrams}. It 
is easily verified that this approximation is exact in the large $N$ 
limit, in the sense that it realizes the resummation of all the 
diagrams that survive this limit.

With this method, one obtains the following  renormalized 
self-consistent gap equation for the mass $m$
\beq\label{GAP2} m^2\,=\,12 g_\mu^2\int\frac{{\rm d}^3k}{(2\pi)^3}\,
\frac{n(\varepsilon_k)}{\varepsilon_k}\,+\, \frac{3 g_\mu^2  m^2}{4
\pi^2}\left(\log \frac{m^2}{\bar\mu^2} \,-1\right). \eeq where the
renormalized coupling $g_\mu$ is related to the bare one  $g_B$  by
($\epsilon=4-d$):
\begin{equation}\label{RENL}
\frac{1}{g^2_\mu}=\frac{\mu^{\epsilon}}{g_B^2}+\frac{3}{2\pi^2\epsilon}.
\end{equation}
By using
Eq.~(\ref{RENL}), one can check that the solution $m^2$ of
Eq.~(\ref{GAP2}) is independent of $\mu$ (to all orders in $g$). It 
is easily seen that the $\beta$-function that is deduced from 
Eq.~(\ref{RENL}) is  only one third of the
one-loop $\beta$-function (see Eq.~(\ref{renorm0})). This is a 
peculiar feature of $\Phi$-derivable approximations 
\cite{Blaizot:2003an}: the particular resummation involved in the 
solution of the gap equation corresponds, for the scattering 
amplitude, to the iteration of the basic vertex in a single channel 
out of three. The correct one-loop $\beta$-function is recovered when 
one keeps in the $\Phi$-derivable approximation the skeleton  of 
order $g^4$ \cite{Blaizot:2003an}.

One can also easily compute the 
pressure
\begin{equation}\label{PPHI}
P=-T\int\frac{{\rm
d}^3k}{(2\pi)^3}\,\log(1-{\rm 
e}^{-\beta\varepsilon_k})
+\frac{m^2}{2}\int\!\!\frac{d^3k}{(2\pi)^3}\,\frac{n(\varepsilon_k)}
{2\varepsilon_k}\,+\frac{m^4}{128\pi^2}\,.
\end{equation}
One 
can verify explicitly, but this is obvious from a simple analysis of 
the diagrammatic content of the approximation, that this expression 
for the pressure  is
perturbatively correct to order $g^3$.
Besides, the complete, self-consistent results, as obtained by numerical
evaluation of Eqs.~(\ref{GAP2}) and (\ref{PPHI}), extrapolate
smoothly to large values of $g$. We present the numerical results of
the self-consistent mass and pressure in Sec.~\ref{sec:numerical}
(in Figs.~\ref{fig:mass} and \ref{fig:pressure}) when we compare
them to the results obtained in this paper by RG techniques.

\section{Non-perturbative renormalization group
\label{sec:RGandLPA}}

The basic strategy of the NPRG 
\cite{Ringwald:1989dz,Berges:2000ew,Bervillier:2005za} consists in
adding to the classical action (\ref{eq:classicalS})
a regulator, conveniently defined in momentum space as
\begin{equation}
\Delta S_{\kappa}[\varphi] = \frac{T}{2}\sum_{\omega_{n}} 
\int\frac{d^{d-1}q}{(2\pi)^{d-1}} \varphi_{i}(q) R_{\kappa}(q) 
\varphi_{i}(-q),
\end{equation}
where $q=(q_0=i\omega_{n}, {\bf q})$, with the Matsubara frequencies 
$\omega_{n}=2\pi nT$
and $R_{\kappa}$ is a cutoff function,
depending on the continuous parameter $\kappa$,
whose specific form will be given shortly. The role of $\Delta S_\kappa$
is to suppress the fluctuations with momenta $q\lesssim \kappa$,
while leaving unaffected those with $q \gtrsim \kappa$. Thus typically
$R_\kappa(q)\rightarrow \kappa^2$ when $q \ll \kappa$ and
$R_\kappa(q)\rightarrow 0$ when $q \gg \kappa$.
There is a large freedom in the choice of $R_\kappa(q)$, abundantly discussed
in the literature 
\cite{Ball:1994ji,Comellas:1997ep,Liao:1999sh,Litim:2000ci,Litim:2001up,Polchinski:1983gv,Litim:2001fd,Canet:2002gs}.

In this paper, we use a regulator which depends only  on the 
spatial components of the momenta, ${\bf q}$,
but not on the energy variable $q_{0}$. It is chosen of the form
\begin{eqnarray}
R_{\kappa}^{{\rm th}}({\bf q}) & = & (\kappa^{2}-{\bf 
q}^{2})\theta(\kappa^{2}-{\bf q}^{2}).\label{eq:LitimIppVacTh}
\end{eqnarray}
A regulator that preserves Euclidean
invariance, and is of the form of Eq.~(\ref{eq:LitimIppVacTh}) with 
${\bf q}\longrightarrow q$,
could of course be
used \cite{Litim:2000ci}. However, such a regulator leads to several difficulties: since
the regulator cuts off frequencies sharply, the contribution of
single Matsubara frequencies become visible as we vary $\kappa$,
resulting in an oscillatory behavior of the flows of various
quantities. Furthermore, because the fluctuations are not entirely
suppressed at the microscopic scale $\Lambda$ by such a regulator, 
initial conditions for quantities
like the pressure need to be fine-tuned with temperature dependent
counterterms. Although this can be done, it leads to  unnecessary 
complications.

The NPRG constructs a family of
effective actions $\Gamma_\kappa [\phi]$ (with $\phi$ the
expectation value of the field), in which the magnitude of long
wavelength fluctuations are controlled by the regulator $R_\kappa$.
The effective action $\Gamma_\kappa [\phi]$ interpolates between the
classical action obtained for $\kappa = \Lambda$ (with $\Lambda$ the
microscopic scale at which the fluctuations are essentially
suppressed) and the full effective action obtained when all
fluctuations are taken into account, that is, when
$\kappa\rightarrow 0$. One can write for $\Gamma_\kappa[\phi]$ an
exact flow equation
\cite{Tetradis:1993ts,Morris:1993qb,Morris:1994ie}
\begin{equation}
\partial_{\kappa}\Gamma_{\kappa}[\phi]=\frac{T}{2}{\rm 
tr}\sum_{\omega_{n}}\int\frac{d^{d-1}{\bf 
q}}{(2\pi)^{d-1}}\partial_{\kappa}R_{\kappa}({\bf 
q})\left[\Gamma_{\kappa}^{(2)}+R_{\kappa}\right]_{q,-q}^{-1}\,,\label{eq:flowequationGamma}\end{equation}
where $\Gamma^{(2)}_\kappa$ is the second derivative of 
$\Gamma_\kappa$ with respect to $\phi$.

\begin{figure}
\begin{center}\includegraphics[%
   scale=0.8]{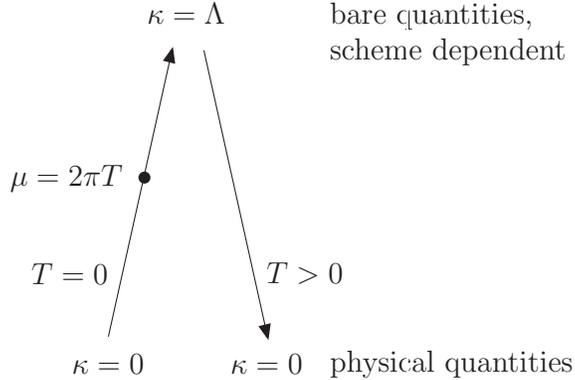}\end{center}
\caption{Basic concept of how to apply the flow equation
at finite temperature.
The coupling $g$ is extracted from the vacuum flow at the scale 
$\kappa = \mu$ with $\mu = 2\pi T$.
\label{cap:concept}}
\end{figure}

Conceptually, we will follow the strategy developed in 
Ref.~\cite{Tetradis:1992xd} and apply the flow equations in the way
illustrated in Fig.~\ref{cap:concept}: Starting
with given physical parameters at $\kappa=0$, we integrate the flow
equations up from $\kappa=0$ to $\Lambda$ thereby removing quantum
fluctuations step by step in order to arrive at bare quantities at a
chosen scale $\Lambda$. If $\Lambda$ is chosen big enough, only
renormalizable  parameters survive and the
system can be described by a simple set of bare parameters. Starting
from these bare parameters we can follow the flow down from
$\kappa=\Lambda$ to 0, but this time with the temperature $T$ turned on.
The physical quantities are then obtained at
$\kappa=0$.

Note however that the coupling constant is fixed at the scale
$\kappa=2\pi T$ on the $T=0$ flow, as discussed above.  As for the
mass at scale $\Lambda$, it is adjusted so that the mass at
$\kappa=0$ vanishes. Note that this procedure induces a specific
scheme dependence attached to the choice of the regulator. We should
keep in mind this scheme dependence when comparing with results of
perturbation theory or those of the 2PI resummation, that are
obtained in another scheme.

Computationally, it might be numerically hard to integrate the flow
equations up from $\kappa=0$ to $\Lambda$. But this is just equivalent
to integrating them down, while carefully adjusting the bare parameters
such that the flow will arrive at physically desired quantities at
the end of the flow at $\kappa=0$.

\subsection{Local potential approximation}

A commonly used approximation to solve the flow equation for
$\Gamma_\kappa[\phi]$ at zero external momenta is the local potential approximation
(LPA). In this approximation, one assumes that
the effective action has the form \cite{Morris:1994ki,Berges:2000ew}
\begin{equation}
\Gamma_{\kappa}^{{\rm LPA}}[\phi]=\int_{0}^{\beta}d\tau\int 
d^{d-1}x\left\{ 
\frac{1}{2}\partial_{\mu}\phi_{i}\partial_{\mu}\phi_{i}+V_{\kappa}(\rho)\right\} 
,
\end{equation}
where $\rho\equiv\phi_{i}\phi_{i}/2$ and $V_{\kappa}(\rho)$ is the
effective potential. The propagator
of the scalar field can be decomposed into its transverse ($G_{T}$) 
and longitudinal
($G_{L}$) components:
   \begin{equation}
   G_{ij}(\kappa;q) = 
G_{T}(\kappa;q)\left(\delta_{ij}-\frac{\phi_{i}\phi_{j}}{2\rho}\right)
   + G_{L}(\kappa;q)\frac{\phi_{i}\phi_{j}}{2\rho}\,.
   \end{equation}
The equation for the potential, derived from Eq.~(\ref{eq:flowequationGamma})
by assuming $\phi$ to be constant, then reads
\begin{equation}
\partial_{\kappa}V_{\kappa}(\rho)=\frac{T}{2}\sum_{\omega_{n}}\int\frac{d^{d-1}{\bf 
q}}{(2\pi)^{d-1}}\left[\partial_{\kappa}R_{\kappa}({\bf 
q})\right]\left\{ 
\left(N-1\right)G_{T}(\kappa;q)+G_{L}(\kappa;q)\right\}, 
\label{eq:basicpotential}
\end{equation}
with
\begin{eqnarray}
G_{T}(\kappa;q) & = & 
\frac{1}{q^{2}+V'_\kappa(\rho)+R_{\kappa}(q)}\,, \label{eq:GTRk}\\
G_{L}(\kappa;q) & = & \frac{1}{q^{2}+V'_\kappa(\rho)+2\rho 
V''_\kappa(\rho)+R_{\kappa}(q)}\,,\label{eq:GLRk}
\end{eqnarray}
with $V'_\kappa(\rho)=dV_\kappa/d\rho$ and 
$V''_\kappa(\rho)=d^{2}V_\kappa/d\rho^{2}$.
The LPA may be viewed as the leading order
in a systematic expansion of the effective action in powers
of the derivatives of the field. Such an expansion has been shown to 
exhibit quick apparent convergence if the
regulator is appropriately chosen 
\cite{Litim:2001up,Canet:2002gs,Canet:2003qd,Bagnuls:2000ae}. 

The choice (\ref{eq:LitimIppVacTh}) of the regulator leads to the 
following flow
equation for the potential (see App.~\ref{app:Flowequation})
\begin{eqnarray}
\partial_{\kappa}V_{\kappa}(\rho)
   & = & K_{d-1}\kappa^{d}
   \left\{ \left(N-1\right)
     \frac{2 n(\omega_\kappa^T) + 1}{2\omega_\kappa^T}
   + \frac{2 n(\omega_\kappa^L) + 1}{2\omega_\kappa^L}
     \right\} ,
\label{eq:potentialTstart}
\end{eqnarray}
where we have defined
\begin{eqnarray}
\omega_\kappa^T & \equiv & \sqrt{V'_\kappa(\rho)+\kappa^{2}} , \nonumber\\
\omega_\kappa^L & \equiv & \sqrt{V'_\kappa(\rho)+2\rho
V''_\kappa(\rho)+\kappa^{2}} \,.\label{eq:omegakappaTL}
\end{eqnarray}
Here, $n(q_{0})=(\exp(q_{0}/T)-1)^{-1}$ is the bosonic distribution
function
  and
$K_{d}=S_{d}/(d(2\pi)^{d})$ with $S_{d}=2\pi^{d/2}/\Gamma(d/2)$. In
the cases that we shall study below, either the second ($N=1$ scalar theory) or
the first term (large $N$ limit) in the braces will contribute to the
flow. Note that, because of the one-loop structure of the flow
equation,  the r.h.s.~of Eq.~(\ref{eq:potentialTstart})  naturally
separates into a ``thermal contribution'', which involves the
statistical factors and which vanishes when $T\rightarrow 0$, and a
``vacuum contribution'' that contains no statistical factors.

\subsection{Truncation and elementary analysis of the flow equations}

\label{sec:truncation}
\begin{figure}[t!]
\begin{center}
\includegraphics*[scale=0.8,angle=0]{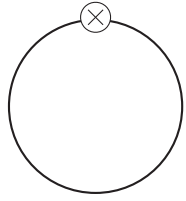}
\end{center}
\caption{ Diagrammatic illustration of the r.h.s.~of the  flow equation
of the effective action, Eq.~(\ref{eq:diffset1a}).
The crossed circle represents
an insertion of the regulator $\partial_\kappa R_\kappa$, and the 
thick line a full
propagator
with mass $m_\kappa$.
\label{fig:mastereq}}
\end{figure}

In the next section, we shall present numerical solutions of the
flow equation, Eq.~(\ref{eq:potentialTstart}). However, much
insight can be gained by considering a simplified version of this
equation, obtained by expanding the potential $V(\rho)$ around
$\rho = 0$ so that
$m_\kappa^2=V_\kappa '(\rho)|_{\rho=0}$ 
and $g_\kappa^2 = V_\kappa ''(\rho)|_{\rho=0} /8$.
We
shall give only the formulae for the case $N=1$,
but they easily generalize to arbitrary $N$. We start from the flow
equation (\ref{eq:potentialTstart})
and insert a truncated potential of the form
\begin{eqnarray}
V_{\kappa}(\phi) & = & 
V_{\kappa}+\frac{m_{\kappa}^{2}}{2}\phi^{2}+g_{\kappa}^{2}\phi^{4}+h_{\kappa}^{2}\phi^{6}+\dots 
\nonumber \\
  & = & 
V_{\kappa}+m_{\kappa}^{2}\rho+4g_{\kappa}^{2}\rho^{2}+8h_{\kappa}^{2}\rho^{3}+\dots
\label{eq:PotentialExpansion}
\end{eqnarray}
Truncating the series at order $O(\rho^{3})$, and neglecting
$h_{\kappa}$ and higher order coefficients, result in replacing
$V'_\kappa(\rho)+2\rho V''_\kappa(\rho)+\kappa^{2}$ by
$\kappa^{2}+m_{\kappa}^{2}+24g_{\kappa}^2\rho$. We obtain the
following set of coupled differential equations at orders
$\rho^{0}$, $\rho$, and $\rho^{2}$:
\begin{eqnarray}
O(\rho^{0}):\quad\partial_{\kappa}V_{\kappa} & = & 
K_{d-1}\kappa^{d}\frac{1+2n\left(\epsilon_{\kappa}\right)}{2\epsilon_{\kappa}}\,,\label{eq:diffset1a}\\
O(\rho^{1}):\quad\partial_{\kappa}m_{\kappa}^{2} & = & 
-6g_{\kappa}^{2}K_{d-1}\kappa^{d}\frac{1+2n\left(\epsilon_{\kappa}\right)-2\epsilon_{\kappa}n'(\epsilon_{\kappa})}{\epsilon_{\kappa}^{3}}\,,\label{eq:diffset1b}\\
O(\rho^{2}):\quad\partial_{\kappa}g_{\kappa}^{2} & = & 
27g_{\kappa}^{4}K_{d-1}\kappa^{d}\frac{1+2n\left(\epsilon_{\kappa}\right)-2\epsilon_{\kappa}n'(\epsilon_{\kappa})+\frac{2}{3}\epsilon_{\kappa}^{2}n''(\epsilon_{\kappa})}{\epsilon_{\kappa}^{5}}\,\label{eq:diffset1c}
\end{eqnarray}
with the short-hand notation 
$\epsilon_{\kappa}:=\sqrt{\kappa^{2}+m_{\kappa}^{2}}$.
The right hand side of Eq.~(\ref{eq:diffset1c}) would change if
one wanted to include the coupling $h_{\kappa}$ (of order 
$g_\kappa^6$), but Eqs.~(\ref{eq:diffset1a})
and (\ref{eq:diffset1b}) would not.
These equations
have a simple interpretation in terms of the diagrams depicted in
Figs.~\ref{fig:mastereq}, \ref{fig:dGamma2}, and \ref{fig:4point}:
these diagrams can be easily calculated with the formulae given in
Appendix A for the finite temperature loop integrals.

\begin{figure}[t!]
\begin{center}
\includegraphics*[scale=0.8,angle=0]{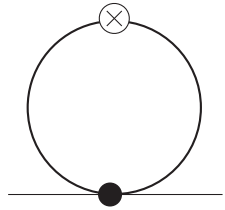}
\end{center}
\caption{\label{fig:dGamma2} Diagrammatic illustration of the r.h.s.~of 
the flow equation for the  2-point function, e.g.~the mass, 
Eq.~(\ref{eq:diffset1b}). The black dot
denotes the four-point function (the coupling $g_\kappa$) and the 
thick line the full propagator $G_\kappa$
(including the regulator, see Eqs.~(\ref{eq:GTRk}), (\ref{eq:GLRk})). 
The circled cross represents the insertion of
$\partial_\kappa R_\kappa$.}
\end{figure}

\begin{figure}[t!]
\begin{center}
\includegraphics*[scale=0.8,angle=0]{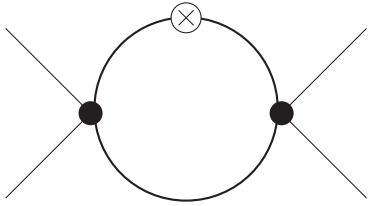}
\end{center}
\caption{\label{fig:dGamma4} Diagrammatic illustration of the r.h.s.~of 
the flow
equation for the 4-point function, 
showing the contribution to the flow of the coupling $g_\kappa$ 
included in Eq.~(\ref{eq:diffset1c})
(the contribution of the 6--point function is ignored in this equation).
Black disks represent the coupling $g_\kappa$.
The three channels 
are included in (\ref{eq:diffset1c}) by the proper factor 3
(external momenta vanish). The crossed circle represents an
insertion of $\partial_\kappa R_\kappa$, and the thick line a full
propagator.\label{fig:4point}}
\end{figure}

The general character of the flow can easily be inferred from a
simple analysis of Eqs.~(\ref{eq:diffset1a}), (\ref{eq:diffset1b})
and (\ref{eq:diffset1c}). For instance, the  coupling follows a
four-dimensional flow from $\Lambda$ to $\kappa\approx2\pi T$, then a
three dimensional flow before
it freezes when $\kappa \simeq m_D$. This can be seen from
Eq.~(\ref{eq:diffset1c}): At large values of $\kappa$, the bosonic
distribution function can be neglected and $\epsilon_\kappa \approx
\kappa$ so that $\partial_\kappa g_\kappa^2 \propto
g_\kappa^4/\kappa$ and we recover the characteristic logarithmic
flow in $d=4$ dimensions. In the opposite limit $\kappa\rightarrow
0$, the value of $\epsilon_\kappa$ is limited by the thermal mass
$\epsilon_\kappa \approx m_D \sim g T$ (see Eq.~(\ref{eq:pert14})), 
and the flow
becomes  $\partial_\kappa g_\kappa^2 \propto g_\kappa^4\kappa^4$,
leading  quickly  to a constant value for $g_\kappa^2$. In the
intermediate range when we still have $\kappa \gtrsim m_\kappa$ and
thus $\epsilon_\kappa \approx \kappa$, but one can already expand
the thermal distribution function for $\kappa \lesssim T$ (see
Eq.~(\ref{eq:nbcombinationexpanded}) below), one finds
$\partial_\kappa g_\kappa^2 \propto g_\kappa^4/ \kappa^2$, which is
compatible with a three dimensional flow, with $g_\kappa\sim \kappa$
at small $\kappa$. This qualitative behavior will be confirmed
by the numerical results presented in the next section.

\section{Numerical results\label{sec:numerical}}

The numerical integration of the flow equation 
(\ref{eq:potentialTstart}) (using the Runge-Kutta
method with adaptive step-size) is hampered by the flow of the 
potential over several orders of
magnitude. In order to keep the zero-temperature flow under control, 
one can introduce
dimension-less variables \cite{Berges:2000ew}, but these will not
be suitable for the thermal flow, which freezes in dimensionful 
variables (but would diverge
in dimensionless variables as $\kappa \rightarrow 0$). However, one 
can take advantage of the
fact that a large part of the flow cancels between thermal and vacuum 
contributions, as will be
described in subsection \ref{subsec:pressure}, and obtain useful 
results for the potential, and
therefore the pressure, mass, and coupling.
In the first part of this section, we present results for the scalar 
theory with $N=1$ component, and in the last subsection for the large 
$N$ limit where the
LPA becomes exact \cite{D'Attanasio:1997ej,Tetradis:1995br}.

\subsection{Coupling}

\begin{figure}
\begin{center}\includegraphics[%
   scale=0.8]{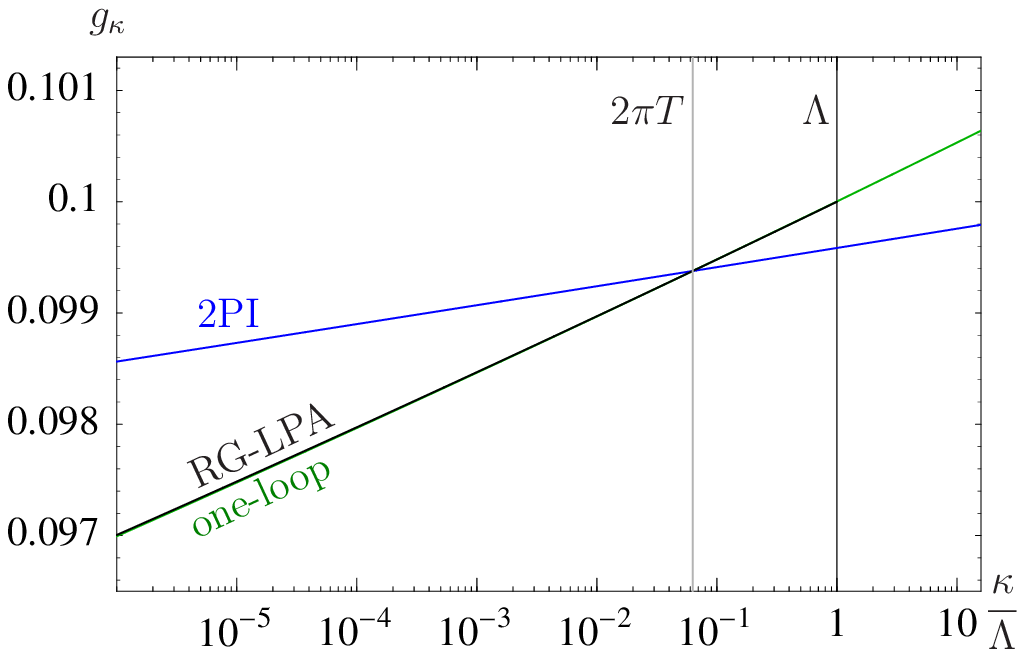}~~\includegraphics[%
   scale=0.8]{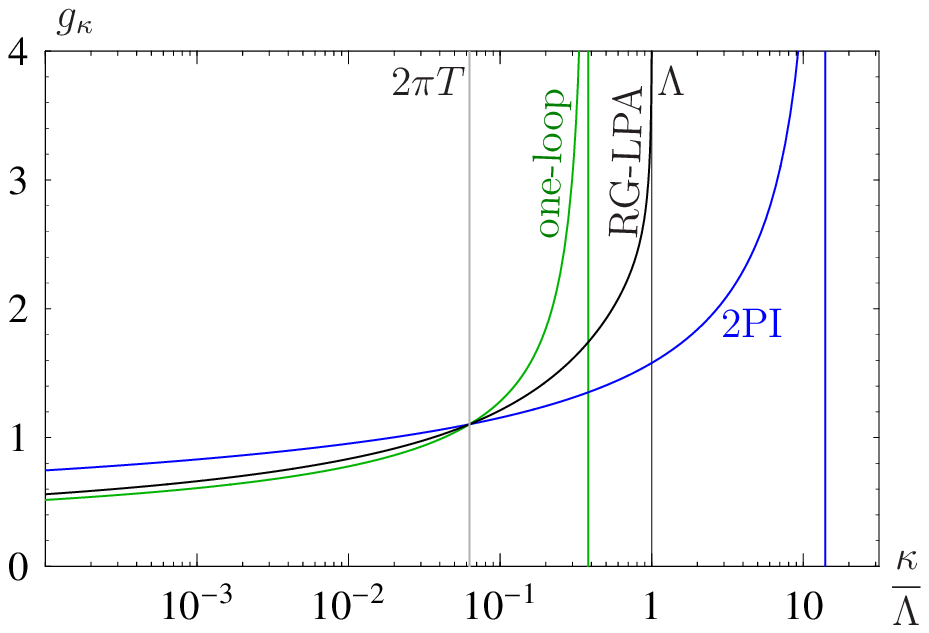}\end{center}
\caption{Comparison of the flow of the coupling at zero temperature for
$g\equiv g(2\pi T) = 0.0994$ ($g_{\Lambda}=0.1$)
and for
$g = 1.1032$ ($g_{\Lambda}=4$). At weak coupling one observes that the 2PI
$\beta$-function differs form the one-loop result by a factor $1/3$. At
larger coupling the Landau pole is observed in the running of the
coupling.\label{fig:couplingzeroT}}
\end{figure}

\begin{figure}
\begin{center}\includegraphics[%
   scale=0.8]{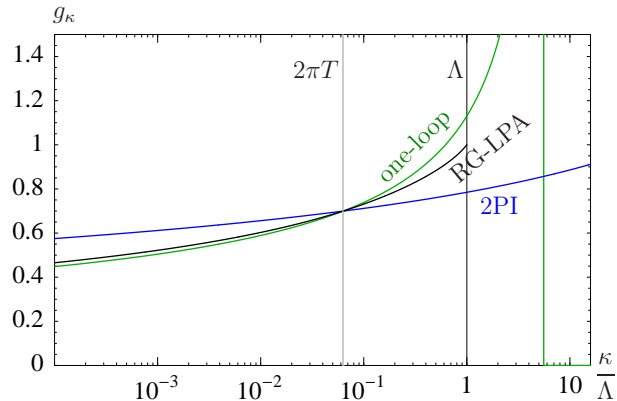}\end{center}
\caption{
  Same as Fig.~\ref{fig:couplingzeroT} for
$g = 0.699$ ($g_\Lambda=1$). 
\label{fig:couplingzeroTg1}
}
\end{figure}

Figures~\ref{fig:couplingzeroT} and \ref{fig:couplingzeroTg1} show
the LPA flow of the coupling $g_\kappa$ at zero temperature, for 
different values
of the $g=g(2\pi T)$, comparing with one-loop results as well as with
those of the self-consistent approximation described in Sec.~\ref{sec:pert_theory}. The 
three plots correspond to $T = \Lambda/100$, and the numerical 
correspondence between $g$ and
  bare coupling $g_\Lambda$ is indicated in the caption.
By construction, the curves cross each other at the scale $\kappa = 2\pi T$.
As shown by the left panel of
Fig.~\ref{fig:couplingzeroT}  the LPA $\beta$-function
agrees with  the one-loop $\beta$-function,
Eq.~(\ref{eq:betafunction}), at weak coupling, reflecting the scheme
independence of its first coefficients. Also visible on
Fig.~\ref{fig:couplingzeroT} is the fact, already mentioned,  that
the 2PI resummation scheme leads to a $\beta$-function that is
smaller by a factor $3$. As discussed in section \ref{sec:towardssc},
at larger couplings (right panel of
Fig.~\ref{fig:couplingzeroT}), all three curves exhibits  the presence of
the Landau pole where $g_\kappa$ diverges. Note that the perturbative
estimate of the position of the Landau pole in
Eq.~(\ref{eq:LandauPole}) is exceeded by the NPRG
flow, and the moderate 2PI
flow predicts a Landau pole that lies  further away.  The one-loop 
calculation in Fig.~\ref{fig:couplingzeroT}
(right panel) does not make sense for the large value of the
coupling since then, as seen in the right panel of 
Fig.~\ref{fig:couplingzeroT}, $\Lambda_{\rm L}<\Lambda$. Figure
\ref{fig:couplingzeroTg1} shows the flow at some intermediate
coupling $g(2\pi T) = 0.699$ ($g_\Lambda = 1$).

\label{page:discussionrange}
Thus,  the presence of the Landau pole limits the applicability of
the NPRG  to not too large couplings, as already discussed in
Sec.~\ref{sec:towardssc}. There is however another limitation 
that comes from the fact that  the denominator in 
(\ref{eq:potentialTstart}),
given by (\ref{eq:omegakappaTL}), becomes imaginary.
This is to be expected to happen when the coupling gets too large, as 
can be seen from the perturbative
estimate: $V_\Lambda '(0) = m_\Lambda^2 \approx -g_\Lambda^2 
\Lambda^2 /(2\pi^2)$
(see Eq.~(\ref{eq:diffset3b}) below).

\begin{figure}
\begin{center}\includegraphics[%
   scale=0.8]{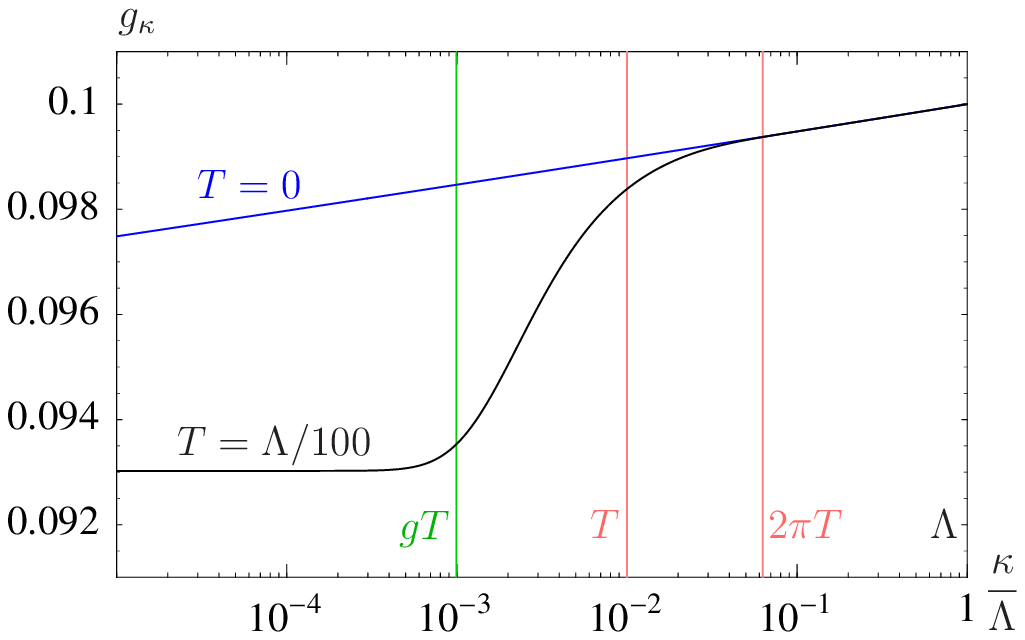}~~\includegraphics[%
   scale=0.8]{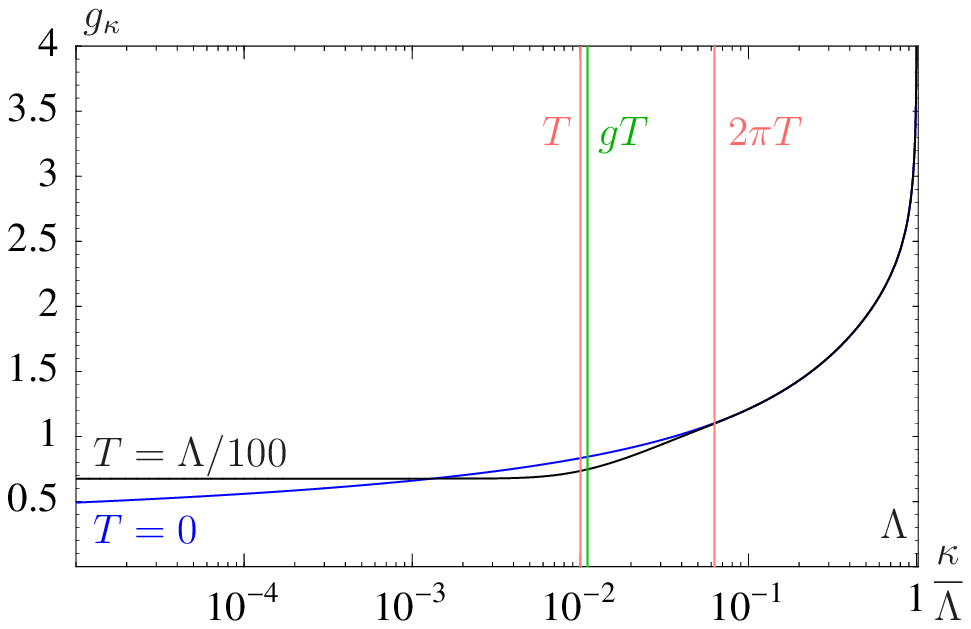}\end{center}
\caption{Flow of the coupling at finite temperature for $g=0.0994$ 
($g_{\Lambda}=0.1$), left, and for
$g=1.1032$ ($g_{\Lambda}=4$),
right.
For the smaller coupling, the value of $g_\kappa$ drops only moderately from 
$\kappa = \Lambda$ to $2\pi T$ and 
the thermal coupling reaches a fixed value at $g_{\kappa=0} = 0.093$. 
For the larger coupling, the beta function forces a steep drop from 
$g_{\Lambda}=4$ to $g = 1.1032$, 
and further to $g_{\kappa=0} = 0.68$ for the thermal coupling. 
\label{fig:flowcoupling}}
\end{figure}

\begin{figure}
\begin{center}\includegraphics[%
   scale=0.8]{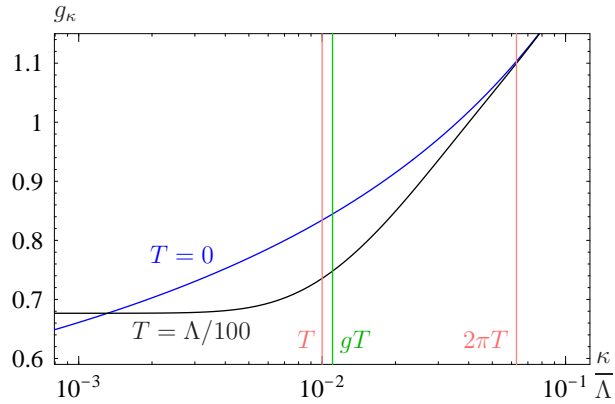}\end{center}
\caption{Enlarged view of Fig.~\ref{fig:flowcoupling} (right) to show 
the region of dimensionally reduced flow.
\label{fig:flowcouplingenlarged}}
\end{figure}

In Fig.~\ref{fig:flowcoupling} we turn on the temperature, and
compare the vacuum flow with the flow obtained at a temperature $T =
\Lambda / 100$. At weak coupling (left panel of
Fig.~\ref{fig:flowcoupling}), the scales $g T$ and $T$ are well
separated, and we observe the general features discussed in
Sec.~\ref{sec:truncation}: From $\kappa = \Lambda$ down to $2\pi T$,
the vacuum ($T=0$) flow of the coupling and the thermal flow agree.
While the vacuum flow continues to decrease logarithmically all the
way down to $\kappa = 0$, below $\kappa = 2\pi T$ the thermal flow
starts to deviate from the vacuum flow in a dimensionally reduced
manner. The three-dimensional flow stops at the scale $\kappa \sim g
T$, where the thermal mass is generated.
The situation does not change qualitatively for larger values of the coupling
in the right panel of Fig.~\ref{fig:flowcoupling} (enlarged in Fig.~\ref{fig:flowcouplingenlarged}): The main 
difference is that the
range of the three-dimensional flow shrinks as $g T$ approaches $T$. 
We shall return to this at the end of the
next subsection.

\subsection{Mass}

\begin{figure}
\begin{center}\includegraphics[%
   scale=0.8]{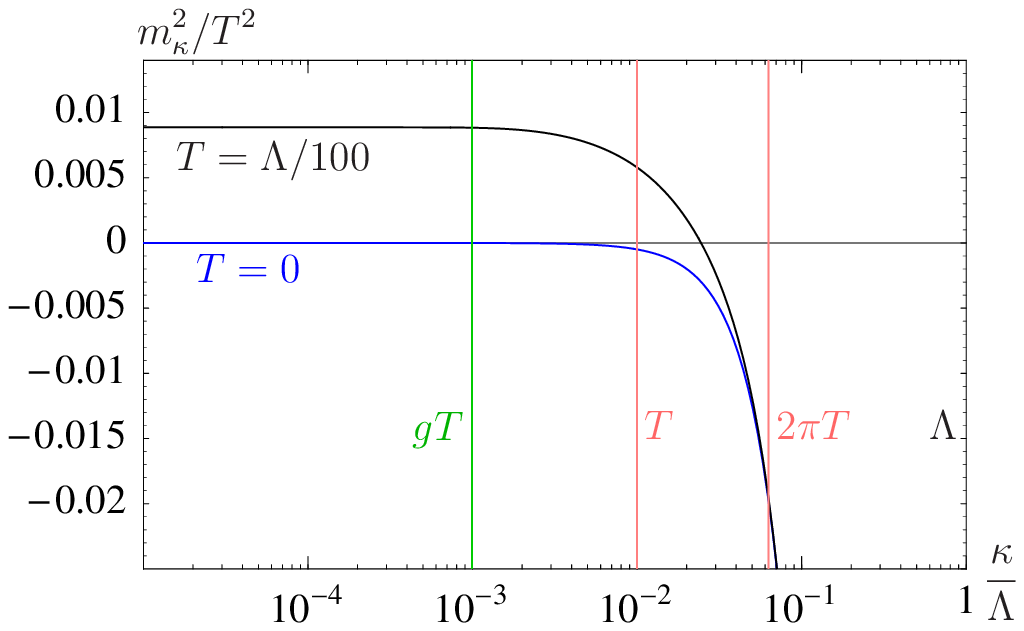}~~\includegraphics[%
   scale=0.8]{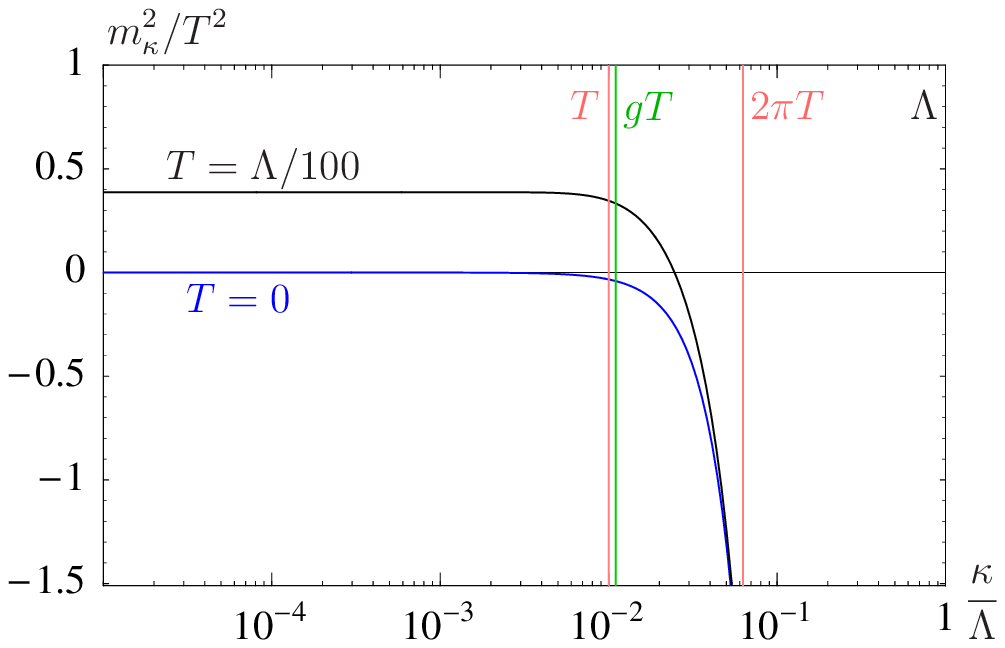}\end{center}
\caption{Flow of the mass for $g=0.0994$ ($g_{\Lambda}=0.1$), left, 
and for $g=1.1032$
($g_{\Lambda}=4$), right. The mass reaches the value $m = 0.094 T$ 
for $g= 0.0994$
and
$m = 0.622 T$ 
for $g= 1.1032$
as $\kappa \rightarrow 0$.
(For $T=0$, the quantity plotted is $m_{{\rm vac},\kappa}^2/T^2$.)
\label{fig:flowmass}}
\end{figure}

The freezing of the flow at soft scales is caused by the
thermally generated mass $m_\kappa^2=V'_\kappa(\rho=0)$,
shown in Fig.~\ref{fig:flowmass}.
As is the case for the coupling, the flow in the range $2\pi T
\lesssim \kappa \leq \Lambda$ agrees between vacuum and thermal
flow. The system starts at a large negative mass squared, which can 
be estimated perturbatively
as $m_{\Lambda}^{2}  = -g_{\Lambda}^{2} \Lambda^{2}/(2\pi^{2}) +
O(g_\Lambda^4)$, see Eq.~(\ref{eq:diffset3b}). In the course of the
flow, the mass acquires positive contributions through quantum
fluctuations. The initial condition $m_\Lambda^2$ has been tuned at
zero temperature, such that $m_{\kappa=0}^2$ vanishes. At finite
temperature, the flow of the mass deviates below the scale $\kappa =
2\pi T$ and builds up the thermal mass $m \simeq g T$ that can be observed in
Fig.~\ref{fig:flowmass}.

\begin{figure}
\begin{center}\includegraphics[%
   scale=0.8]{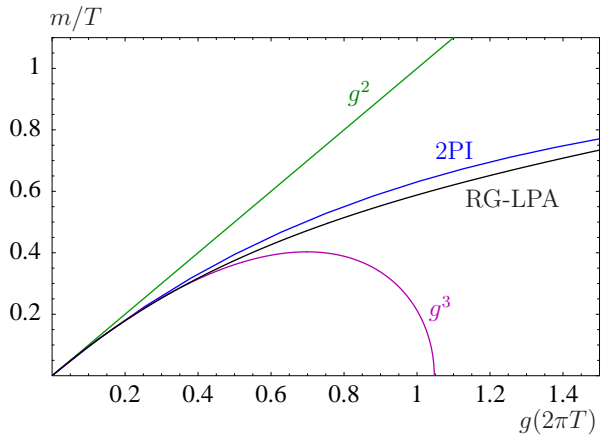}\end{center}
\caption{Thermal mass
for scalar theory with $N=1$ component
as a function of the coupling. The result obtained from
the non-perturbative renormalization group labeled {}``RG-LPA''
gets close to the result obtained by 2PI resummation. Both show moderate
behavior where the perturbative contributions represented by curves
of order $g^{2}$ and $g^{3}$ fail to converge.
\label{fig:mass}}
\end{figure}

The values obtained from the flow equation in 
Figs.~\ref{fig:flowcoupling} and \ref{fig:flowmass}
are combined to produce the
result displayed in Fig.~\ref{fig:mass}. For example, starting the
numerical calculation from $g_\Lambda = 4$, one obtains a
coupling $g = g(2\pi T) = 1.1032$.
At this value of the coupling, the mass takes the
value $m/T = 0.622$ (Fig.~\ref{fig:flowmass}) in the physical limit
  $\kappa \rightarrow 0$.
This information is put into Fig.~\ref{fig:mass}, and the procedure 
is repeated for other values
of the coupling. The curve thus obtained, labeled ``RG-LPA'', is 
smooth, in contrast to
those displaying the results of strict perturbation theory through
orders $g^{2}$ and $g^{3}$. For comparison, in Fig.~\ref{fig:mass}
we also plot the results obtained through the 2PI resummation
described in Sec.~II \cite{Blaizot:2000fc}. The difference obtained
between the RG-LPA and 2PI approaches has two origins: on the one 
hand, the two calculations
correspond to two different $\beta$-functions: $\beta_{\rm 2PI} = \beta_{\rm
one-loop} / 3$ (see Fig.~\ref{fig:couplingzeroT}). On the other hand,
the two calculations use different renormalization schemes: the 2PI curve
shows the scale choice in the $\overline{{\rm MS}}$-scheme with
$\bar{\mu}_{\overline{{\rm MS}}}=2\pi T$. In comparison, the NPRG
approach defines the coupling from the vacuum flow at $\kappa=2\pi
T$ and the quantities at other scales depend on the choice of the regulator.

From Figs.~\ref{fig:flowcoupling} to \ref{fig:mass} we can already 
extract one of the main results of our work.
Both in the weak and the strong coupling regimes, the scale 
dependence of the running coupling and of the mass
exhibits a similar behavior:
in both cases, thermal effects start to play a role around $\kappa 
\simeq 2\pi T$
and saturate at $\kappa \simeq m_D \simeq gT$. When going from weak 
to strong coupling,
there is a competition between two effects: on the one side, the 
range of the values of $\kappa$
where thermal effects are important shrinks because the thermal 
mass increases with the coupling;
on the other side, the amplitude of the three-dimensional flow 
increases with increasing values of $g$.
Both effects are visible in Figs.~\ref{fig:flowcoupling} and 
\ref{fig:flowcouplingenlarged}. The net effect
is that the variation of the coupling along the three 
dimensional flow is limited; for instance,
in Fig.~\ref{fig:flowcouplingenlarged} for $g=1.1032$, it amounts to 
a 30\% effect.

One can also observe in Fig.~\ref{fig:flowcouplingenlarged} that, at 
strong coupling, the flow saturates at
a value of $\kappa \lesssim g T$. This is because the thermal mass 
reaches a value lower than $g T$,
as illustrated in Fig.~\ref{fig:mass}. In fact, one may argue that 
the thermal mass cannot exceed a value
$\simeq 2\pi T$, since otherwise all thermal effects would disappear, which 
is in contradiction with the existence
of such thermal mass.

\subsection{Pressure}
\label{subsec:pressure}
\begin{figure}
\begin{center}\includegraphics[%
   scale=0.8]{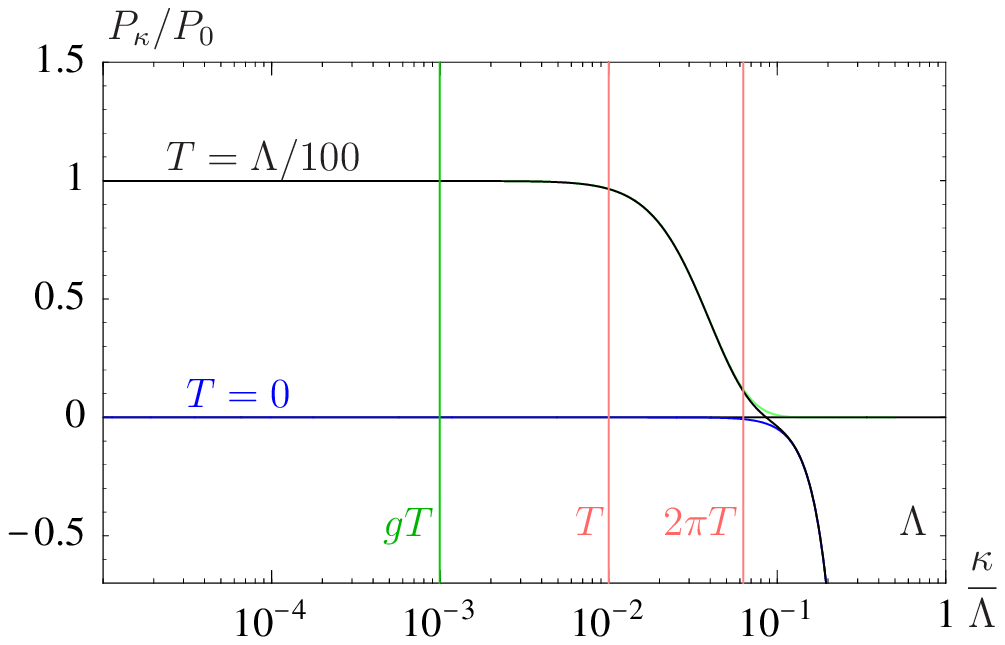}~~\includegraphics[%
   scale=0.8]{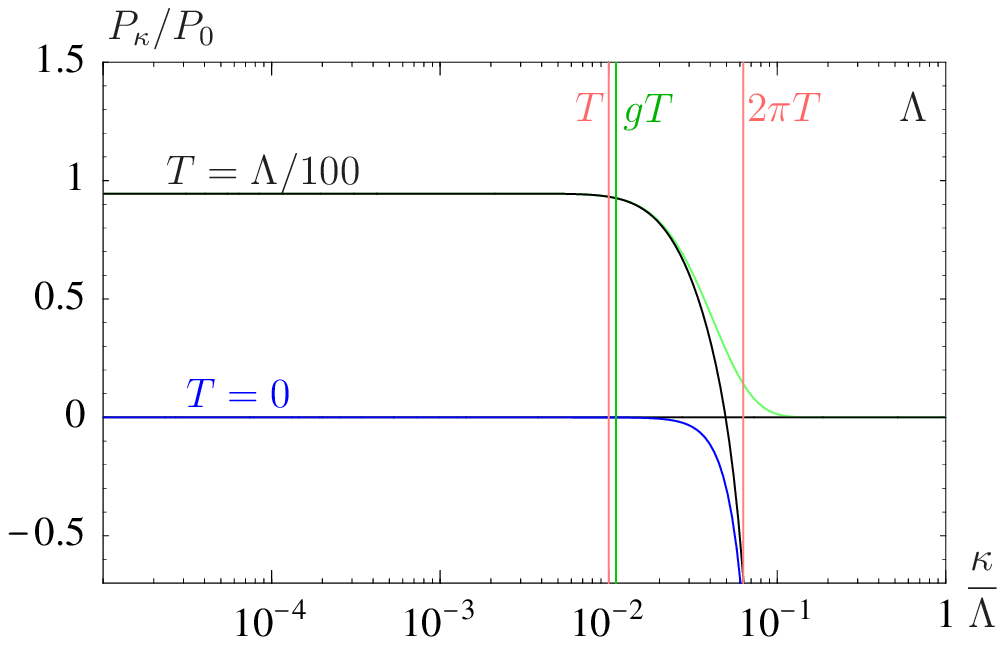}\end{center}
\caption{Comparison of the flow of the pressure in the weak-coupling regime
for $g=0.0994$ ($g_{\Lambda}=0.1$), left panel, and in the 
strong-coupling regime
for $g=1.1032$ ($g_{\Lambda}=4$), right panel, with $T=\Lambda/100$.
The additive constant is chosen such that the vacuum 
pressure vanishes
as $\kappa \rightarrow 0$. The thermal pressure reaches the value of
$P=0.9984 P_0$ for $g=0.0994$ 
and
$P=0.945 P_0$ for $g = 1.1032$ 
as $\kappa \rightarrow 0$.
\label{fig:flowpressure}}
\end{figure}

Figure \ref{fig:flowpressure} shows the flow of the pressure $P = -V(\rho = 0)$ for weak 
and strong couplings, normalized to the free pressure $P_0 = (\pi^2/90)T^4$.
We choose $V_\Lambda(\rho = 0)$ such that the 
zero-temperature
pressure $P_{\kappa = 0}(T=0)$ vanishes. This requires however a delicate cancellation. 
Indeed, a perturbative calculation
(see Eq. (\ref{eq:diffset3a}) below) shows that the vacuum pressure 
at $\kappa=\Lambda$
is of the order $\Lambda^{4}$.
Thus, for example, both curves
for the $g=1.1032$ plot start at $P_{\Lambda}=-415,470.518 P_0$
and acquire a relative difference during the flow from $\kappa=\Lambda$
to $\kappa=0$ of $0.945 P_0$. Clearly, this cancellation over six
orders of magnitude poses a challenge to the numerics. One should
note though that the thermal flow above $\kappa\gtrsim2\pi T$
lies exponentially close to the zero temperature flow due to the bosonic
distribution function in Eq.~(\ref{eq:potentialTstart}), and the
bulk of the thermal contribution builds up between $\kappa\simeq 2\pi 
T$ and $\kappa \simeq g T$.
This difference between thermal and vacuum flows is shown as
a light 
curve in the plots. It is therefore sufficient to start
the integration of the thermal pressure at some 
intermediate scale $2\pi T\ll\Lambda_{1}\ll\Lambda$.
The zero-temperature and finite-temperature flows of the potential 
only reach a maximum value of the order of
$\Lambda_{1}^{4}\lll\Lambda^{4}$ and their difference is numerically 
much easier to handle.
On the other hand, $\Lambda_1$ can not be chosen too small, or contributions
to the result will be neglected.
Practically, a value of $\Lambda_1 = 20 T$ turned out to be a good
compromise between
loss of accuracy
due to neglecting $\kappa \geq \Lambda_1$ contributions and
gain of accuracy due to reduced numerical cancellation.
This procedure also improves the numerical accuracy of the thermal mass,
shown in Fig.~\ref{fig:flowmass}, although the situation is not as 
critical there
(the mass only grows as $\kappa^2$, as opposed to the pressure that 
grows as $\kappa^4$).

Since the pressure, mass, and coupling constant all approach constant 
values as $\kappa \to 0$,
dimensionless variables that one may introduce for the zero-temperature flow
cease to be useful:
For example, as the dimensionful potential $V_{\kappa\rightarrow0}$
approaches a constant, the dimensionless variable diverges as 
$v_{\kappa\rightarrow0}\sim1/\kappa^{d}$.
Numerically, we have still followed the approach of dimensionless variables
also for thermal quantities, as $\kappa$ does not need to become
too small to measure the limiting value reached when $\kappa\rightarrow0$.

\begin{figure}
\begin{center}\includegraphics[%
   scale=0.8]{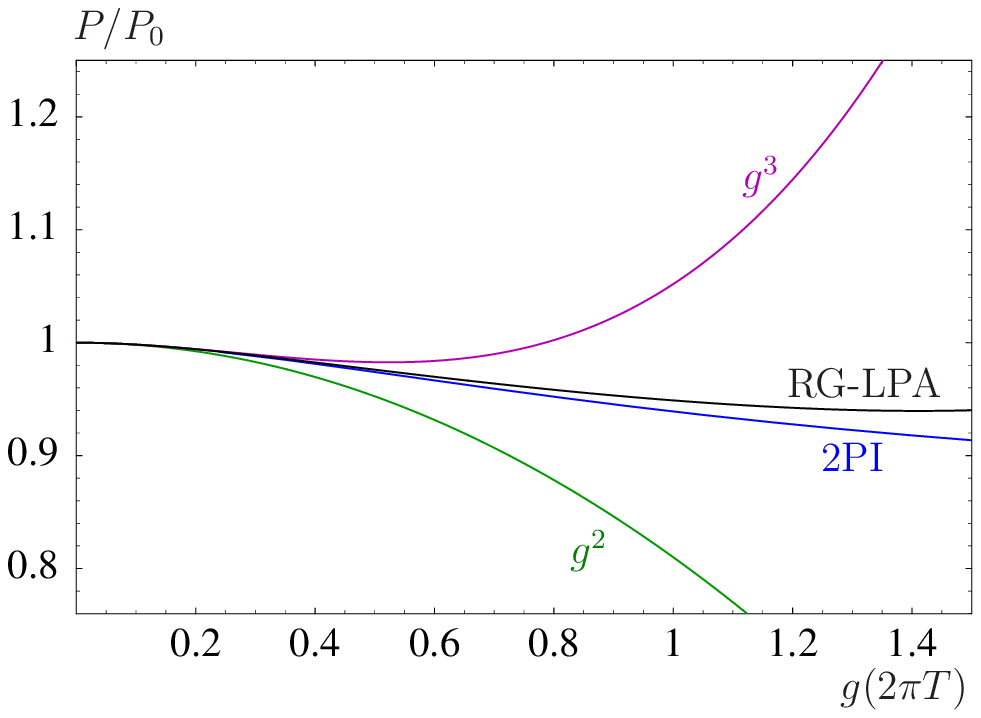}~~\includegraphics[%
   scale=0.8]{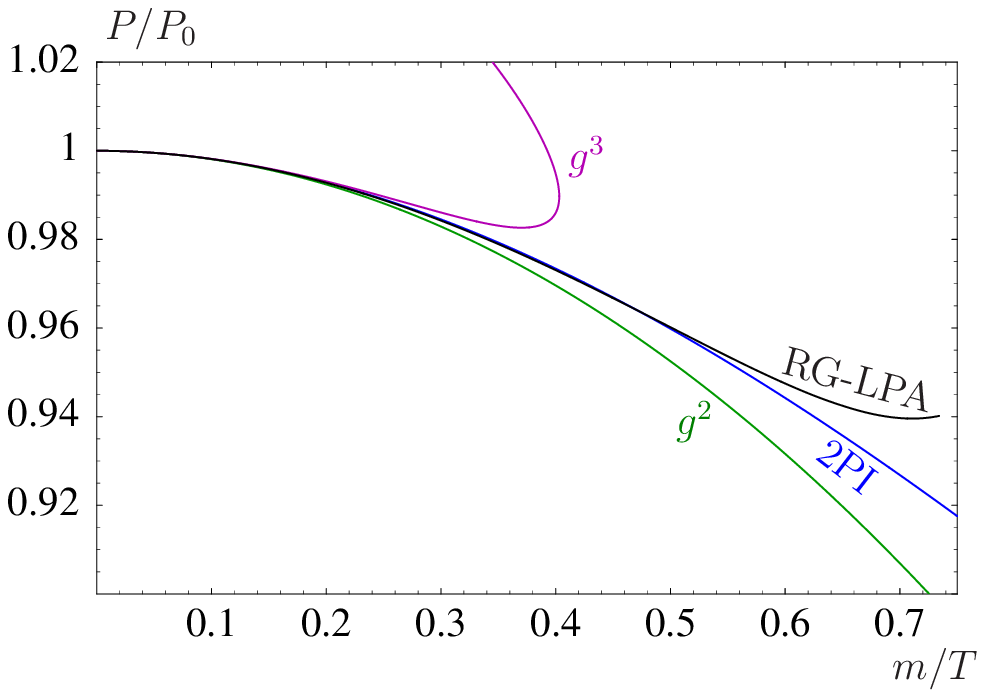}\end{center}
\caption{Pressure  as a function of the coupling (left panel) and as
a function of the mass (right panel). As two physical,
scale-independent quantities are compared, the right plot is
renormalization scale independent and shows an improved agreement
between RG-LPA and 2PI results. \label{fig:pressure}
\label{fig:pressuremass}}
\end{figure}

The left panel of Fig.~\ref{fig:pressure} shows the pressure as a function of
$g$. It follows combining results from Figs.~\ref{fig:flowcoupling}
and \ref{fig:flowpressure}, in the same  way as Figs.~\ref{fig:flowcoupling}
and \ref{fig:flowmass} lead to Fig.~\ref{fig:mass}.
Due to the limitations of the maximum value of the coupling that we can use
(see discussion on page \pageref{page:discussionrange}),
it is necessary to reduce the ratio $T/\Lambda$
for the larger couplings in Fig.~\ref{fig:pressure}.
The range depicted (up to $g\approx 1.5$) is about the range
in which both pressure and thermal mass are still independent
of this choice of $T/\Lambda$, up to plot accuracy.
For the largest $g \sim 1.5$ we have chosen $T/\Lambda = 1/20$;
this agrees for smaller $g$ with $T/\Lambda = 1/50$ or $1/100$.
The value $T/\Lambda = 1/10$ could extend the NPRG method to even larger $g$,
but then the result visibly starts to deviate from the $T/\Lambda = 
1/20$ result.
Similar to the plot for the mass in Fig.~\ref{fig:mass},
we observe an improved result ``RG-LPA'' over perturbation theory 
through orders $g^2$
and $g^3$. We obtain a deviation of similar magnitude when compared 
to the 2PI resummation scheme,
part of which is caused as mentioned above by different 
renormalization schemes.
Note that the 2PI curve lies below the RG-LPA curve.

We can circumvent the problem of renormalization scale by comparing
only physical quantities. Figure~\ref{fig:pressuremass} (right panel) displays
the pressure as a function of the mass.
For the LPA curve, both quantities, pressure and mass, are
extracted at $\kappa = 0$, so there is no more reference in this plot to a
particular choice of a scale for the coupling $g$.
The perturbative contributions $g^2$ and $g^3$ on the other hand
can be considered as scale-independent only up to the order they have 
been calculated to.
These curves may change when extracted at a different scale.
In this plot we observe improved agreement between the RG-LPA curve
and the 2PI curve, but also note that they do not completely agree.
Particularly, at larger $m/T$, the LPA  curve starts to bend up again.
This is the region where the NPRG equations get sensitive to the Landau pole
and may also fail because of imaginary parts in the denominator of 
the flow equation
as discussed on page \pageref{page:discussionrange}, so this region should
be considered with care (the fact that the 2PI curve does not show the
same bending as the RG-LPA one presumably reflects the fact that 
the 2PI Landau pole is
artificially pushed to higher values because
of the factor $1/3$ in the 2PI $\beta$-function).
The break-down of perturbation theory can be pinpointed where the
$g^{3}$ contribution deviates visibly. This curve bends backwards,
because $m(g)$ is not a monotonous function at this order. 
Surprisingly, the $g^{2}$
curve seems to remain a good approximation to the non-perturbative approaches.

\subsection{Large $N$ limit}

In the large $N$ limit, only the transverse propagator in
Eq.~(\ref{eq:potentialTstart}) contributes to the flow. The various
quantities scale in the following way as $N\rightarrow\infty$:
\begin{eqnarray}
P,\, V,\,\rho & \sim & N\,,\nonumber \\
V'_\kappa(\rho),\, m^{2} & \sim & 1\,,\nonumber \\
g^{2} & \sim & 1/N\,. \label{eq:largeNscaling}
\end{eqnarray}
It is then advantageous to plot results as a
function of an effective coupling $g_{{\rm eff}}
\equiv g\sqrt{N}$. Note that both $P/P_0$ and $m/T$ are $N$-independent
quantities.

\begin{figure}
\begin{center}%
\includegraphics[scale=0.8]{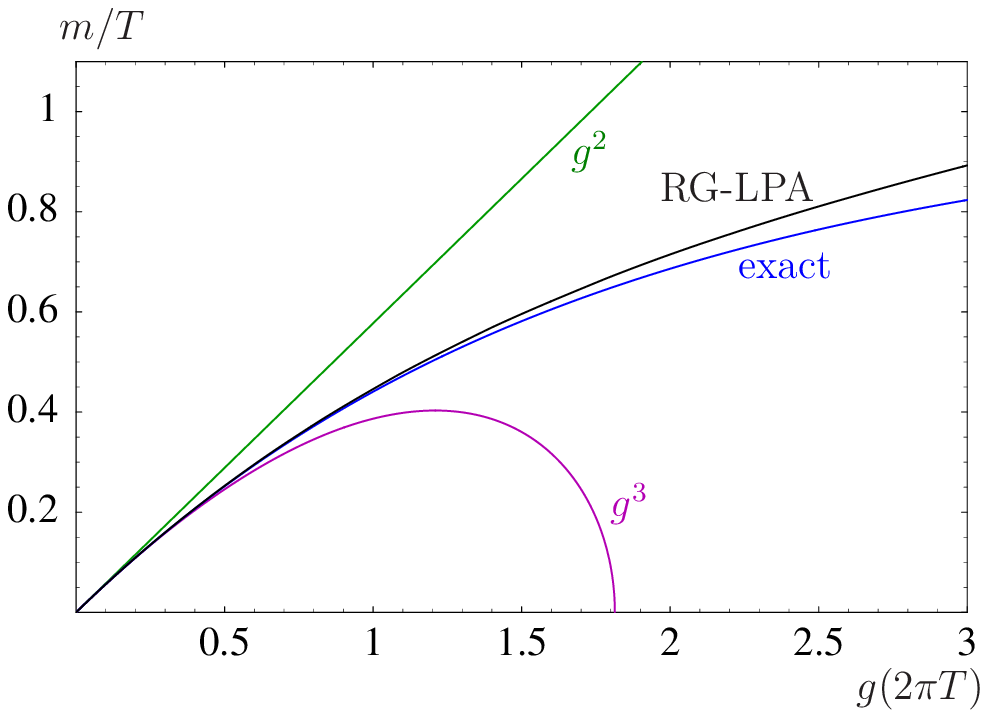}~~%
\includegraphics[scale=0.8]{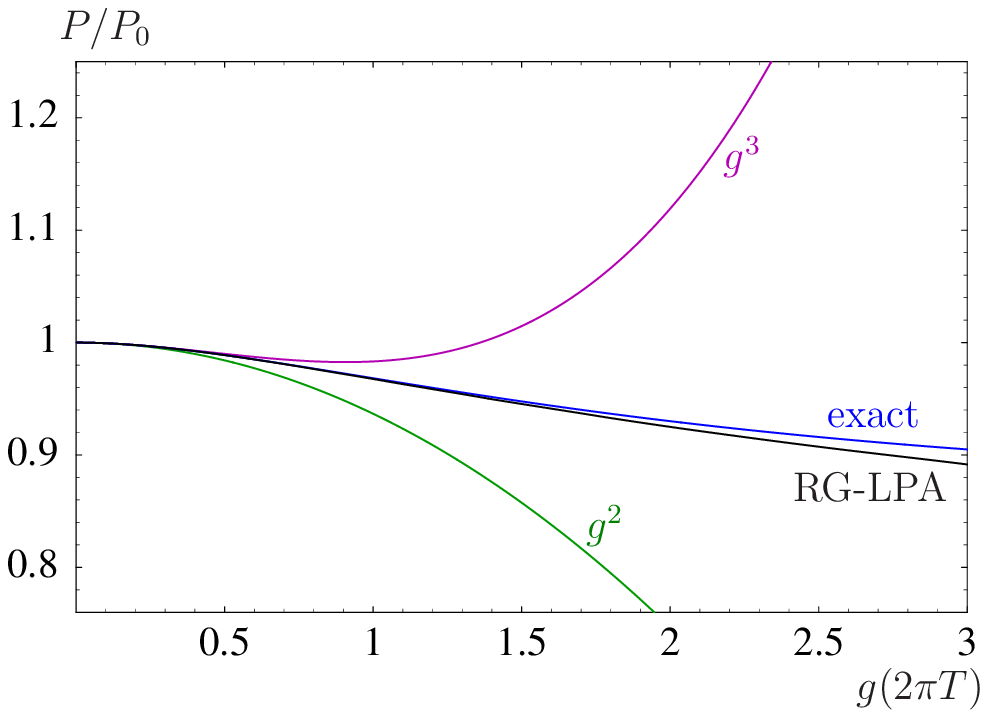}%
\end{center}
\caption{Comparison of mass (left panel) and pressure (right panel) 
at large $N$, normalized to temperature and the free
pressure, respectively. The coupling $g$ denotes the effective 
coupling $g_{{\rm eff}}=g\sqrt{N}$
in the limit $N\rightarrow\infty$.\label{fig:preslargeN}}
\end{figure}

\begin{figure}
\begin{center}\includegraphics[%
   scale=0.8]{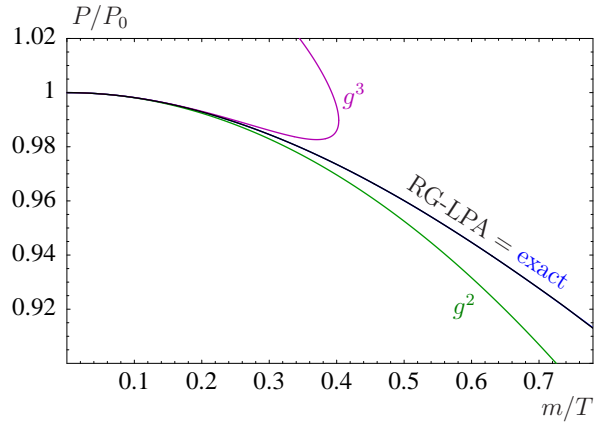}\end{center}
\caption{Pressure as a function of the mass in the large $N$ limit.
In the comparison of scale-independent quantities, the RG-LPA numerical result
exactly reproduces the large $N$ limit.
\label{fig:pressuremasslargeN}}
\end{figure}

In Fig.~\ref{fig:preslargeN} we plot the predictions of the RG-LPA 
for the mass and the pressure
as a function of the coupling constant,
together with the exact large $N$-limit results, which are known (see 
e.g.~\cite{Blaizot:2003tw}).
In the large $N$ limit, the LPA approximation also gets exact
\cite{D'Attanasio:1997ej,Tetradis:1995br}.
The reason why the curves of Fig.~\ref{fig:preslargeN} do not
exactly agree is because
of the already mentioned scheme dependence. If one
plots only scale-independent quantities, like the pressure as a
function of the thermal mass, then one recovers indeed a full
agreement between the RG-LPA and the known
exact large $N$ limit results (see
Fig.~\ref{fig:pressuremasslargeN}).

\section{Perturbative expansion of the flow equation\label{sec:perturbative}}

In the spirit of other resummation schemes proposed in the literature,
we show that the  NPRG in the LPA,
reproduces perturbation theory up to order $g^3$.
Of course, the full numerical evaluation of the flow equation
in 
Sec.~\ref{sec:numerical} includes higher orders in a
non-perturbative way.

We start from the truncated flow equations (\ref{eq:diffset1a}) to
(\ref{eq:diffset1c}) where the neglected term in
(\ref{eq:PotentialExpansion}), $h_\kappa$,  is of order $g_\kappa^6$. 
These equations for the mass (\ref{eq:diffset1b}) and coupling 
constant (\ref{eq:diffset1c}) from a closed set
of equations that can be solved order by order in $g$. The pressure
can then be calculated using the results for $m_\kappa$ and
$g_\kappa$.

In the case of the vacuum flow, the perturbative hierarchy is quite clear:
$m_\kappa^2$ is always $\sim g_\kappa^2\kappa^2$, and one
can therefore expand $\epsilon_\kappa=\sqrt{\kappa^2+m_\kappa^2}\simeq \kappa$
for all $\kappa$.

At finite temperature, on the other hand, a thermal mass of the order $g T$
is generated, and expanding for small $m_\kappa$ necessarily fails 
when $\kappa \lesssim g T$.
In this case, it is useful to introduce an intermediate scale $\xi$
that lies between soft and hard momenta $g T \ll \xi \ll T$ 
(e.g.~$\xi \sim \sqrt{g}\, T$).
Different expansions can then be performed in these two regions:
For hard momenta, $\kappa \gtrsim \xi$,
\begin{equation}
\epsilon_\kappa^2 = \kappa^2 + m_\kappa^2 = \kappa^2 \left( 1 + O( (g 
T / \xi) ^2 ) \right).
\label{eq:largekappaexpansion}
\end{equation}
For soft momenta, $\kappa \lesssim \xi$, we can expand
the thermal distribution function
\begin{eqnarray}
n(\kappa) & = & 
\frac{T}{\kappa}-\frac{1}{2}+\frac{\kappa}{12T}+\frac{T}{\kappa}O( 
(\xi / T)^3 ) .
\label{eq:nIR}
\end{eqnarray}
The leading term, $T/\kappa$, trades one dimension of a 
$d$-dimensional integration by the temperature,
thereby leading to a dimensionally reduced system.
 From the calculation of section \ref{sec:rev-pert}, one expects order 
$g^2$ contributions to
arise from integrals dominated by hard momenta,
and $g^3$ effects to build up at soft momenta.
However, as we shall see, delicate cancellations occur in the course 
of the calculation.
In order to cleanly reproduce strict perturbative results at order $g^3$,
we shall expand the actual running quantities $g_\kappa$ or 
$m_\kappa$ around their leading order
perturbative contributions:
\begin{eqnarray}
g_\kappa^2 & = & g^2 + \delta g_\kappa^2 , \nonumber \\
m_\kappa^2 & = & m^2 + \delta m_\kappa^2 ,
\label{eq:deltagm}
\end{eqnarray}
with $m^2 = g^2 T^2$ (see (\ref{eq:mth2simple})), where $g$ without 
subscript is to be understood, as
before, as $g = g_{{\rm vac}} (\kappa = 2\pi T)$. As we shall see,
in the region $\kappa \geq \xi$ the correction $\delta g_\kappa^2$ 
will be  formally of order $g^4$ (see (\ref{eq:gLambda})),
while for $\kappa \leq \xi$ the correction $\delta g_\kappa^2$ will 
be  formally of order $g^3$ (see (\ref{eq:gth3gIR})) and $\delta 
m_\kappa^2$ of the order $g^2 \xi T$ (see (\ref{eq:mth2gIR}) and 
(\ref{eq:mth3gIR})).
Technical details that amount to $O(g^4)$ or $O(g^4 \ln g)$ 
contributions are deferred to
App.~\ref{app:technicaldetails}; the LPA is not expected to reproduce 
them correctly.

\subsection{Zero-temperature flow}

At zero temperature, all the distribution functions $n(\epsilon_{\kappa})$
and their derivatives vanish in Eqs.~(\ref{eq:diffset1a}) to 
(\ref{eq:diffset1c}), and we are left with the following set
of differential equations, that we expand in powers of $g_{\kappa}^{2}$,
\begin{eqnarray}
\partial_{\kappa}V_{\kappa} & = & 
\frac{K_{d-1}\kappa^{d}}{2\epsilon_{\kappa}}\,=\, 
K_{d-1}\left(\frac{\kappa^{d-1}}{2}-\frac{m_{\kappa}^{2}\kappa^{d-3}}{4}\right)+O(g_{\kappa}^{4}),\label{eq:diffset2a}\\
\partial_{\kappa}m_{\kappa}^{2} & = & 
-\frac{6g_{\kappa}^{2}K_{d-1}\kappa^{d}}{\epsilon_{\kappa}^{3}}\,=\,-6g_{\kappa}^{2}K_{d-1}\kappa^{d-3}+O(g_{\kappa}^{4})\,,\label{eq:diffset2b}\\
\partial_{\kappa}g_{\kappa}^{2} & = & 
\frac{27g_{\kappa}^{4}K_{d-1}\kappa^{d}}{\epsilon_{\kappa}^{5}}\,=\,27g_{\kappa}^{4}K_{d-1}\kappa^{d-5}+O(g_{\kappa}^{6})\,.\label{eq:diffset2c}
\end{eqnarray}
In these equations, we anticipated that the mass $m_{\kappa}^{2}$
will be of order $g_{\kappa}^{2}$ (i.e.~we assume that
   $m_{\kappa=0}$ vanishes, so that $m_{\kappa}^{2}$ is entirely
due to quantum fluctuations). For $d=4$ the solutions of these equations
read:
\begin{eqnarray}
V_{\kappa} & = & 
V_{\Lambda}+\frac{\kappa^{4}-\Lambda^{4}}{48\pi^{2}}+O(g_{\Lambda}^{2}),\label{eq:diffset3a}\\
m_{\kappa}^{2} & = &
-g_{\kappa}^{2}\frac{\kappa^{2}
}{2\pi^{2}}+O(g_{\Lambda}^{4})\,,\label{eq:diffset3b}\\
g_{\kappa}^{2} & = & 
\frac{g_{\Lambda}^{2}}{1-9g_{\Lambda}^{2}\log(\kappa/\Lambda)/(2\pi^{2})}
+O(g_{\Lambda}^{6}) \,. \label{eq:diffset3c}
\end{eqnarray}
In deriving these equations, we have assumed that $g_\Lambda \ll 1$, 
and also taken into account
the possible occurrence of large logarithms in getting Eq.~(\ref{eq:diffset3c}). 
To get Eq.~(\ref{eq:diffset3b}), one integrates the r.h.s.~of Eq.~(\ref{eq:diffset2b}) 
by parts using (\ref{eq:diffset2c}).
Finally, the integration constant has been adjusted so that
$m_{\kappa=0}^{2}$ vanishes, i.e.~we
have $m_{\Lambda}^{2}=-g_{\Lambda}^{2}\Lambda^{2}/(2\pi^{2})$.
Using these results, we can actually proceed one order further in the 
potential:
\begin{eqnarray}
V_{\kappa} & = & 
V_{\Lambda}+\frac{\kappa^{4}-\Lambda^{4}}{48\pi^{2}}+\frac{g_{\kappa}^{2}\kappa^{4}-g_{\Lambda}^{2}\Lambda^{4}}{192\pi^{4}}+O(g_{\Lambda}^{4}).\label{eq:diffset3aa}
\end{eqnarray}

\subsection{Coupling at finite temperature}

In order to find the perturbative solution of 
Eq.~(\ref{eq:diffset1c}) we introduce an intermediate scale  $\xi$, 
as discussed above, with   $g T \ll \xi \ll T$. For $\kappa \geq \xi$ 
thermal effects are subleading and $g_\kappa^2$ can be obtained 
directly from (\ref{eq:diffset3c})
\begin{eqnarray}
g_{\kappa}^{2} =g^{2} + \frac{9g^{4}}{2\pi^{2}} \log 
\frac{\kappa}{2\pi T} + O(g^{6}) .
\label{eq:gLambda}
\end{eqnarray}
In the region  $\kappa \leq \xi$ we can expand the distribution
function in Eq.~(\ref{eq:diffset1c}) according to (\ref{eq:nIR}) 
which gives a remarkably stable
expansion (i.e.~all powers from $\epsilon_{\kappa}^{0}$ to 
$\epsilon_{\kappa}^{4}$
drop out)
\begin{equation}
1+2n\left(\epsilon_{\kappa}\right)-2\epsilon_{\kappa}n'(\epsilon_{\kappa})+\frac{2}{3}\epsilon_{\kappa}^{2}n''
(\epsilon_{\kappa})=\frac{16T}{3\epsilon_{\kappa}}+\frac{\epsilon_{\kappa}^{5}}{5670T^{5}}+
O(\frac{\epsilon_{\kappa}^{6}}{T^{6}})\,.
\label{eq:nbcombinationexpanded}
\end{equation}
The flow equation then becomes
\begin{equation}
\partial_{\kappa}g_{\kappa}^{2}=\frac{24g_{\kappa}^{4}}{\pi^{2}}\frac{\kappa^{4} 
T}{\epsilon_{\kappa}^{6}}
\end{equation}
with $\epsilon_{\kappa}^{2}=\kappa^{2}+m_{\kappa}^{2}$. By writing 
$m_\kappa^2=m^2+\delta m_\kappa^2$ (see Eq.~ (\ref{eq:deltagm})), and 
expanding out the $\delta m_\kappa^2$ correction, we obtain for the 
leading contribution
\begin{equation}
\left.-\frac{1}{g_{\kappa}^2}\right|_{\kappa'}^{\xi} = \left.\frac{3T}{\pi^{2}}
\left(
  \frac{3}{m}\arctan\frac{\kappa}{m}
-\frac{5\kappa^{3}+3\kappa m^{2}}{(\kappa^{2}+m^{2})^{2}}
\right)\right|_{\kappa'}^{\xi} + O(\xi/(g T)).
\label{eq:couplingintegral}
\end{equation}
In the limit $\kappa'\rightarrow 0$ and $g T \ll \xi \rightarrow \infty$
and expanding the result in $g$, we recover the result obtained earlier
\begin{equation}
g_{\kappa=0}^{2}=g^{2}-\frac{9}{2\pi}g^{3}+O(g^3 \xi / T)\,.
\label{eq:coupling3}
\end{equation}
(The expression for general $\kappa$ is given in (\ref{eq:gth3gIR})). 
Note that the result is, to order $g^{3}$,
independent of the temperature $T$.
This may seem at first sight surprising.
However, the only possible effect of of the temperature is a change 
of the renormalization scale,
and this only influences the result at order $O(g^4)$.
A careful calculation through order $g^4$ (presented in Appendix 
\ref{app:coupling4})
reveals how the value of the coupling at $\kappa=0$ is connected to 
the coupling at $\kappa \gtrsim T$:
\begin{equation}
g_{\kappa=0}^2 = g_\kappa^2 - \frac{9g_\kappa^3}{2 \pi}
- \frac{9 g_\kappa^4}{2\pi^2} \left( \log\frac{\kappa}{2 \pi T}
   + \gamma - \frac{61}{12} + \frac{32}{3{\pi }^2} \right) + O(g_\kappa^5),
\label{eq:coupling4text}
\end{equation}
where $\gamma$ is Euler's constant.
Note that this result is obtained from perturbatively expanding the 
flow equation
within the LPA.

\begin{figure}
\begin{center}\includegraphics[%
   scale=0.8]{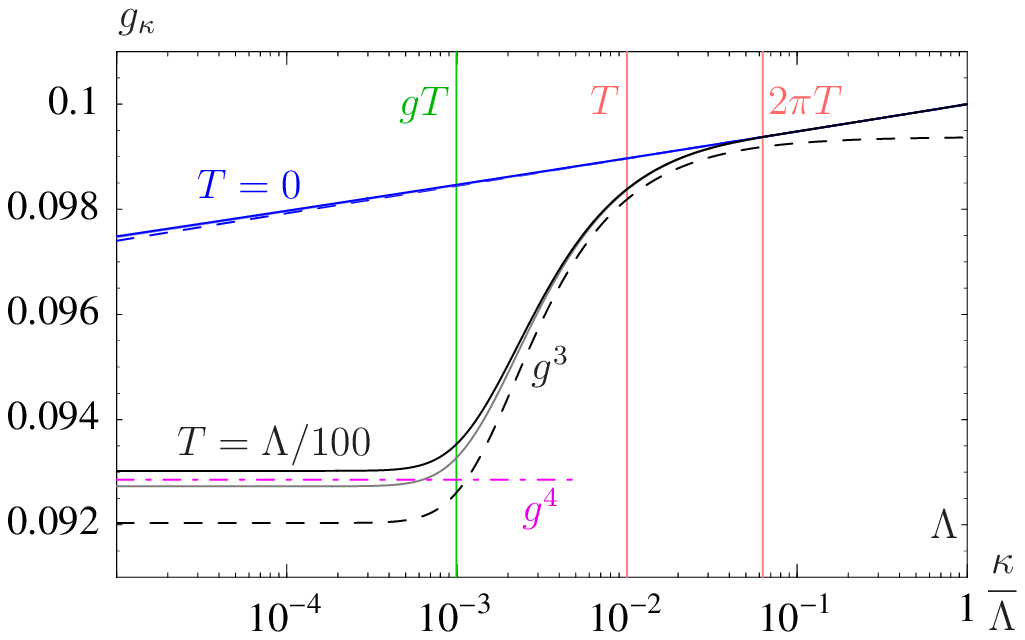}~~\includegraphics[%
   scale=0.8]{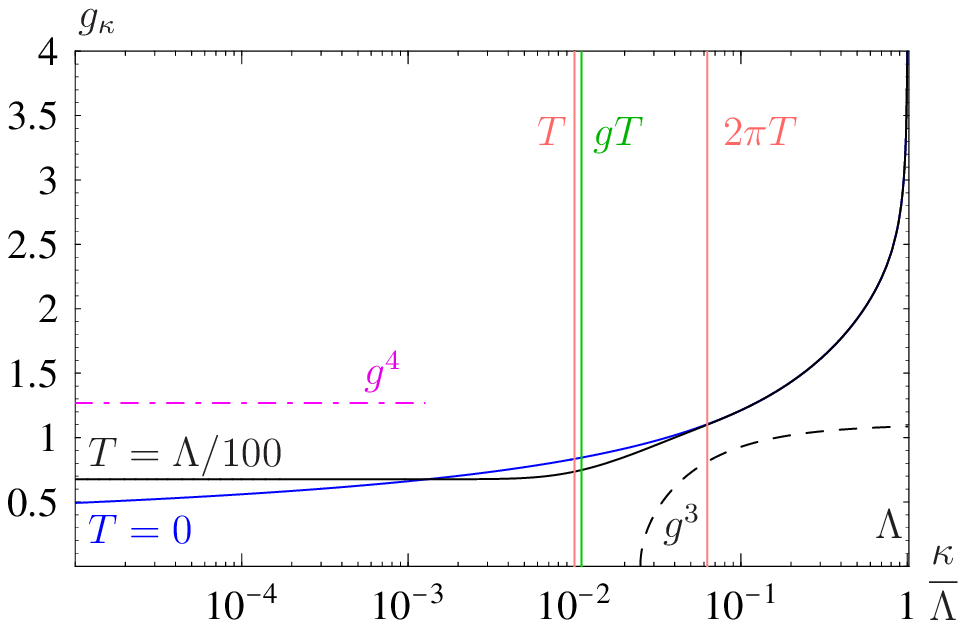}\end{center}
\caption{
Same as Fig.~\ref{fig:flowcoupling}, compared to analytic solutions.
Dashed lines give perturbative solution through order $g^3$ (flowing
with $\kappa$), with $g$ taken at $\kappa = 2\pi T$, and $g^4$ (only
$g_0$ value), with $g$ taken at $\kappa = 2\pi T$.
For the left panel, also the solution of truncated LPA is given by
the light curve. \label{fig:flowcouplingperturbative} }
\end{figure}

Figure \ref{fig:flowcouplingperturbative} shows a comparison with
the numerical solutions. The curves labeled ``$g^3$'' show the sum of
the $g^2$ value at $\kappa=2 \pi T$ and the $\kappa$-dependent
correction of order $g^3$ from Eq.~(\ref{eq:gth3gIR}). This formula
is strictly speaking only valid in a region $\kappa \leq \xi$ and
introduces an error of the order ${\mathcal O} (g T/\xi)$. This
error is the reason why the $g^3$ curve does not agree with the
numerical value of $g$ at $\kappa = 2\pi T$, but reaches this value
only for $\kappa \rightarrow \infty$. Also, through order $g^3$
there is no renormalization scale dependence, and the analytical
solution does not follow the vacuum flow of the numerical result.
The curves labeled ``$g^4$'' in
Fig.~\ref{fig:flowcouplingperturbative} show the result
(\ref{eq:coupling4text})
with 
$g(\kappa = 2\pi T)$. 

For the left plot of Fig.~\ref{fig:flowcouplingperturbative} also
the result of truncated LPA, obtained by numerically integrating
Eqs.~(\ref{eq:diffset1a}) to (\ref{eq:diffset1c}), is given. While for small
coupling in the left plot of Fig.~\ref{fig:flowcouplingperturbative}
the truncated LPA misses the full numerical result of the scattering
amplitude $g_{\kappa=0}$ by 0.3\%, at the larger coupling $g(\Lambda)
= 4$, the truncated LPA would underestimate the value of
$g(\kappa=2\pi T)$ by 28\% (not shown in the plot). 
If we instead
lower the cutoff $\Lambda$ such that the vacuum couplings at $\kappa = 2\pi T$ agree, 
the truncated LPA will give a result for $g_{\kappa=0}$ that is 23\% larger than the full
numerical result. Thus, while this truncation provides reliable 
estimates at small
coupling, it cannot be used at strong coupling.

\subsection{Thermal mass}

We consider now finite temperature effects and focus first on the
thermal mass. This is defined by separating the mass in equation 
(\ref{eq:diffset1b})
into vacuum and thermal pieces:\begin{eqnarray}
m_{{\rm vac},\kappa}^{2} & = & m_{\kappa}^{2}(T=0)\,,\\
m_{{\rm th},\kappa}^{2} & = & m_{\kappa}^{2}(T)-m_{{\rm 
vac},\kappa}^{2}\,.\end{eqnarray}
The thermal contribution can then be calculated using $\epsilon_{\kappa}:=
\sqrt{\kappa^{2}+m_{{\rm th},\kappa}^{2}+m_{{\rm vac},\kappa}^{2}}$,
and reads:
\begin{equation}
m_{{\rm th},\kappa'}^{2}=-6K_{d-1}\int_{\Lambda}^{\kappa'}d\kappa\,\kappa^{d}
\left\{ g_{\kappa}^{2}\frac{1+2n(\epsilon_{\kappa})-2\epsilon_{\kappa}n'
(\epsilon_{\kappa})}{\epsilon_{\kappa}^{3}}-g_{{\rm vac},\kappa}^{2}\frac{1}
{(\kappa^{2}+m_{{\rm vac},\kappa}^{2})^{3/2}}\right\} \,.
\label{eq:mth2}
\end{equation}

\subsubsection{Thermal mass to leading order\label{sec:mth2}}

As we have seen earlier, the $g^2$ contribution to the thermal mass 
is dominated by hard momenta.
Using Eqs. (\ref{eq:largekappaexpansion}) and (\ref{eq:deltagm}),
we obtain as leading contribution
\begin{eqnarray}
m_{{\rm th},\kappa'}^{2} & = & 
12g^{2}K_{d-1}\int_{\kappa'}^{\Lambda}d\kappa\,\kappa^{d-3}
\left(n(\kappa)-\kappa n'(\kappa)\right)+O(g^{4})\,.
\label{eq:mth2g}
\end{eqnarray}
Although, strictly speaking, this expression is valid only for 
$\kappa' \geq \xi$, it is
IR safe, and the error introduced by sending $\kappa '$ to 0 is of 
the order $O(g^2 \xi T)$.
On the UV side, the error
introduced by replacing $\Lambda$ by $\infty$ is exponentially small
(it is of the order $O(T \Lambda e^{-\Lambda/T} )$
for the part with $n$, and $O(\Lambda^2 e^{-\Lambda/T} )$ for $n'$).
We finally obtain for $\kappa' = 0$
\begin{eqnarray}
m_{{\rm th},0}^{2} & = & 
12g^{2}K_{d-1}(d-1)\int_{0}^{\infty}d\kappa\,\kappa^{d-3}n(\kappa)
+ O(g^2 \xi T) + O(\Lambda^2 e^{-\Lambda/T} )\,.
\end{eqnarray}
For $d=4$ dimensions, this gives
\begin{equation}
m_{{\rm th},0}^{2} = m^2 =g^{2}T^{2}\,.
\label{eq:mth2simple}
\end{equation}

\subsubsection{Thermal mass at next-to-leading order}

In order to obtain the next-to-leading order in Eq.~(\ref{eq:mth2}), we need
to treat the small $\kappa$ limit carefully. The resulting 
contribution, of order
$g^3$, is calculated by subtracting the leading $g^{2}$
contribution of Eq.~(\ref{eq:mth2g})
from the unexpanded
Eq.~(\ref{eq:mth2}).\begin{eqnarray}
\left.m_{{\rm th},0}^{2}\right|_{g^{3}}
   & = & 6K_{3}\int_{0}^{\Lambda}d\kappa\,\kappa^{4}
   \left\{ g_\kappa^{2} \frac{1 + 2n(\epsilon_{\kappa}) - 
2\epsilon_{\kappa} n'(\epsilon_{\kappa})}{\epsilon_{\kappa}^{3}}
    - g_{{\rm vac},\kappa}^{2} \frac{1}{(\kappa^{2}+m_{{\rm 
vac},\kappa}^{2})^{3/2}} \right.\nonumber \\
  &  & \left.-2g^{2}\frac{n(\kappa)-\kappa 
n'(\kappa)}{\kappa^{3}}\right\} 
+O(g^{4})\,.\label{eq:mthc}\end{eqnarray}
Note that we keep the $\kappa$-dependence in the first line, as the 
flow of $g_\kappa$ is
affected by $g^3$ corrections (see Eq.~(\ref{eq:coupling3})) that may 
change the mass at order $g^3$.
In fact, a careful calculation reveals that such contributions would 
affect the result only at
order $g^4 \ln g$ (see \ref{app:mass4}).

Since the dominant contribution to the integral (\ref{eq:mthc}) stems
from the infrared region, we split the integration range by an intermediate
scale $\xi$ with $g T\ll\xi \ll T$. For hard contributions
$\kappa\geq\xi$ one can expand $m_{\kappa},\, m_{{\rm vac},\kappa} 
(\ll \kappa)$
in the denominators and in $\epsilon_{\kappa}^{2}=\kappa^{2}+m_{\kappa}^{2}$,
and use $g_{{\rm vac},\kappa}^{2}=g^{2}+O(g^{4})$; one observes then 
that the result
vanishes up to $O(g^{4})$.

For soft contributions $0\leq\kappa\leq\xi$
we can expand the thermal distribution function (\ref{eq:nIR}).
To extract the $g^{3}$ term, it is sufficient to keep the leading
term of (\ref{eq:nIR}) and use coupling and mass only up to order 
$g^2$, i.e.~$g_\kappa \rightarrow g$ and $m_\kappa \rightarrow m = g 
T$. Equation (\ref{eq:mthc}) then simplifies
to
\begin{eqnarray}
\left.m_{{\rm th},0}^{2}\right|_{g^{3}} & = & 
\frac{4g^{2}T}{\pi^2} \int_{0}^{\xi\rightarrow 
\infty}d\kappa\left(\frac{\kappa^{4}}{(\kappa^2 + 
m^2)^2}-1\right)\nonumber \\
  & = & -\frac{3Tm}{\pi}g^{2}\,\,=\,\,-\frac{3T^{2}}{\pi}g^{3}.
\label{eq:mth3}
\end{eqnarray}
We recover the correct result for the correction to the mass at order $g^3$.

We can finally write the perturbative thermal mass as
\begin{equation}
m_{{\rm th},0}^{2}=T^{2}\left[g^{2}-\frac{3}{\pi}g^{3}+O(g^{4}\log g)\right]\,.
\label{eq:mthNLO}
\end{equation}

\subsection{Pressure at finite temperature}

For the thermal pressure we expand equation (\ref{eq:diffset1a})
in a similar way, subtracting the vacuum piece
\begin{equation}
p_{{\rm th}}
=-V_{{\rm th},0}=-(V_{\kappa}(T)-V_{\kappa}(T=0))|_{\kappa=0} ,
\end{equation}
and obtain from Eq. (\ref{eq:diffset1a})
\begin{equation}
p_{{\rm 
th}}=K_{d-1}\int_{0}^{\Lambda}d\kappa\,\kappa^{d}\left(\frac{1+2n(\epsilon_{\kappa})}{2\epsilon_{\kappa}}-\frac{1}{2\sqrt{\kappa^{2}+m_{{\rm 
vac},\kappa}^{2}}}\right)\label{eq:presexact}
\end{equation}
where as before $\epsilon_{\kappa}^{2}:=\kappa^{2}+m_{{\rm 
th},\kappa}^{2}+m_{{\rm vac},\kappa}^{2}=\kappa^{2}+m_{\kappa}^{2}$.
The leading contributions at order $g^0$ and $g^2$ come from the 
region $\kappa \geq \xi$,
in which both, $m_{{\rm th},\kappa}^{2}$ and $m_{{\rm vac},\kappa}^{2}$,
can be expanded from $\epsilon_\kappa$ (see 
(\ref{eq:largekappaexpansion})). We get
\begin{eqnarray}
p_{{\rm th}} & = & K_{d-1}\int_{\xi}^{\Lambda}d\kappa\left\{ 
\kappa^{d-1}n(\kappa)\right.\label{eq:presexpand1a}\\
  & - & \frac{\kappa^{d-3}}{4}\left[m_{{\rm 
th},\kappa}^{2}+2\left(m_{{\rm th},\kappa}^{2}+m_{{\rm 
vac},\kappa}^{2}\right)\left(n(\kappa)-kn'(\kappa)\right)\right]\label{eq:presexpand1b}\\
  & + & \left.O((g T/\xi)^{4})\right\}
  \end{eqnarray}
where the first line is of order $g^{0}$ and the second line is of order
$g^{2}$. The third line in this expansion scheme
is formally of order $g^{4}$, but we anticipate that soft modes with 
$\kappa\leq \xi$ will contribute
a term of order $g^{3}$.
The expressions (\ref{eq:presexpand1a}) and (\ref{eq:presexpand1b}) 
are IR safe, and $\xi$ can be sent to 0 in these expressions without 
affecting
the perturbative result through order $g^2$.

\subsubsection{Free pressure}

The free pressure is easily evaluated from the first line of Eq. 
(\ref{eq:presexpand1a})\begin{eqnarray}
p & = & 
\lim_{\Lambda\rightarrow\infty}K_{d-1}\int_{0}^{\Lambda}d\kappa\,\kappa^{d-1}n(\kappa)\nonumber 
\\
  & = & \frac{\Gamma(d/2)}{\pi^{d/2}}{\rm Li}_{d}(1)T^{d}\end{eqnarray}
where ${\rm Li}_{d}(z)$ is the polylogarithm function. This gives
the known values for the free pressure in 2, 3, or 4 dimensions\begin{equation}
p_{d=2,3,4}=\frac{\pi}{6}T^{2},\,\frac{\zeta(3)}{2\pi}T^{3},\,\frac{\pi^{2}}{90}T^{4}\,.\end{equation}

\subsubsection{Pressure at next-to-leading order}

The contribution to order $g^{2}$ is already harder to obtain.
This is because, as Eq.~(\ref{eq:presexpand1b})
shows, both, $m_{{\rm th},\kappa}^{2}$
and $m_{{\rm vac},\kappa}^{2}$, contribute on equal basis to the
$g^{2}$ result of the pressure. Of course, at the end of the flow,
the zero-temperature mass vanishes $m_{{\rm vac},\kappa}^{2}=0$,
as we have tuned the mass $m_{\Lambda}^{2}$ just in this way, but
this does not mean that we can neglect the effect of $m_{{\rm vac},\kappa}^{2}$
during the evolution of the potential from $\kappa=\Lambda$ to $0$,
even if the final result turns out to be strictly proportional to
the thermal mass $p|_{g^{2}}\propto m_{{\rm th}}^{2}$.

We shall evaluate Eq.~(\ref{eq:presexpand1b}) in $d=4$ dimensions
in three pieces $\left.p_{{\rm th}}\right|_{g^{2}}=p_{{\rm 
th}}^{({\rm th\times vac})}+p_{{\rm th}}^{({\rm vac\times 
th})}+p_{{\rm th}}^{({\rm th\times th})}$
with
\begin{eqnarray}
p_{{\rm th}}^{({\rm th\times vac})} & = & 
-\frac{K_{3}}{4}\int_{0}^{\Lambda}d\kappa\,\kappa \, m_{{\rm 
th},\kappa}^{2} \, ,\\
p_{{\rm th}}^{({\rm vac\times th})} & = & 
-\frac{K_{3}}{2}\int_{0}^{\Lambda}d\kappa\,\kappa \,m_{{\rm 
vac},\kappa}^{2} \left(n(\kappa)-kn'(\kappa)\right)  ,\\
p_{{\rm th}}^{({\rm th\times th})} & = & 
-\frac{K_{3}}{2}\int_{0}^{\Lambda}d\kappa\,\kappa \,m_{{\rm 
th},\kappa}^{2} \left(n(\kappa)-kn'(\kappa)\right) .
\end{eqnarray}
For the first term one uses the explicit leading-order expression (\ref{eq:mth2g}) 
for $m_{\rm th}^2$:
\begin{equation}
p_{{\rm th}}^{({\rm th\times 
vac})}=-3g^{2}K_{3}^{2}\int_{0}^{\Lambda}d\kappa\, 
\kappa\int_{\kappa}^{\Lambda}d\kappa'\kappa'\left(n(\kappa')-\kappa'n'(\kappa')\right)\,.
\end{equation}
  One can integrate this expression by parts, obtaining for 
$\Lambda\rightarrow\infty$\begin{eqnarray}
p_{{\rm th}}^{({\rm th\times vac})} & = & 
-\frac{3g^{2}K_{3}^{2}}{2}\int_{0}^{\infty}d\kappa\left(\kappa^{3}n(\kappa)-\kappa^{4}n'(\kappa)\right)\,\,=\,\,-\frac{15g^{2}K_{3}^{2}}{2}\int_{0}^{\infty}d\kappa\,\kappa^{3}n(\kappa)\nonumber 
\\
  & = & -\frac{g^{2}T^{4}}{72}\,.\end{eqnarray}
  This piece is exactly canceled by the following piece of $p_{{\rm th}}$
\begin{equation}
p_{{\rm th}}^{({\rm vac\times 
th})}=-\frac{K_{3}}{2}\int_{0}^{\Lambda}d\kappa\, \kappa\,  m_{{\rm 
vac}}^{2}(n(\kappa)-\kappa n'(\kappa))=+\frac{g^{2}T^{4}}{72}\,,
\end{equation}
where we have used Eq.~(\ref{eq:diffset3b}) for the mass $m_{{\rm 
vac}}^{2}=-g^{2}\kappa^{2}/(2\pi^{2})$,
again in the limit $\Lambda\rightarrow\infty$.

This cancellation reflects a general property. In a perturbative 
calculation, the terms mixing vacuum and thermal
parts may contain ultraviolet divergences coming from vacuum 
subdiagrams. Since there can be no divergent terms
involving the temperature in the final results, such terms must 
cancel \cite{Blaizot:2003an,Blaizot:2004bg,Blaizot:2002nh}.

The residual contribution to $p_{{\rm th}}$ therefore solely comes
from
\begin{equation}
p_{{\rm th}}^{({\rm th\times 
th})}=-\frac{K_{3}}{2}\int_{0}^{\Lambda}d\kappa\, \kappa\, m_{{\rm 
th}}^{2}(n(\kappa)-\kappa n'(\kappa))\,.
\end{equation}
  Identifying parts of the integrand with the flow equation (\ref{eq:diffset1b})
for the thermal mass\begin{equation}
\partial_{\kappa}m_{{\rm 
th}}^{2}=-12g^{2}K_{3}\kappa(n(\kappa)-\kappa 
n'(\kappa))\,,\end{equation}
  we can write the final contribution as
\begin{equation}
\left.p_{{\rm th}}\right|_{g^{2}}=p_{{\rm th}}^{({\rm th\times 
th})}=\frac{1}{24g^{2}}\int_{0}^{\Lambda}d\kappa\, m_{{\rm 
th}}^{2}\partial_{\kappa}m_{{\rm th}}^{2}=-\frac{1}{48g^{2}}m_{{\rm 
th}}^{4}=-\frac{g^{2}T^{4}}{48}\,.
\end{equation}

\subsubsection{Plasmon contribution to the pressure}

In order to calculate the $g^3$ contribution, we proceed analogously to
the calculation of the mass at order $g^{3}$: We form an IR dominated
integral by subtracting the known contributions at order $g^{0}$
and $g^{2}$ in Eqs.~(\ref{eq:presexpand1a}) and (\ref{eq:presexpand1b})
from the exact integral Eq.~(\ref{eq:presexact}):
\begin{eqnarray}
\left.p_{{\rm th}}\right|_{g^{3}} & = & 
K_{d-1}\int_{0}^{\Lambda}d\kappa\,\left\{ 
\kappa^{d}\left(\frac{1+2n(\epsilon_{\kappa})}{2\epsilon_{\kappa}}-\frac{1}{2\sqrt{\kappa^{2}+m_{{\rm 
vac},\kappa}^{2}}}\right)\right.\nonumber \\
  &  &-\kappa^{d-1}n(\kappa) + 
\left.\frac{\kappa^{d-3}}{4}\left(m_{{\rm 
th},\kappa}^{2}+2\left(m_{{\rm th},\kappa}^{2}+m_{{\rm 
vac},\kappa}^{2}\right)\left(n(\kappa)-kn'(\kappa)\right)\right)\right\} 
\,.
  \end{eqnarray}
This integral is dominated by small values $0\leq\kappa<\xi$
and we can replace the upper integration
limit $\Lambda$ by $\xi$ without changing the result at order $g^3$.
We proceed as before, expanding
the distribution function as in Eq.~(\ref{eq:nIR})
and neglecting $m_{{\rm vac},\kappa}^{2}\approx O(g^{2}\kappa^{2})$, 
and obtain
\begin{eqnarray}
\left.p_{{\rm th}}\right|_{g^{3}} & = & T 
K_{3}\int_{0}^{\xi}d\kappa\left\{ 
\frac{\kappa^4}{\epsilon_{\kappa}^{2}}-\kappa^2 + m_{{\rm th}, 
\kappa}^2 \right\} \nonumber \\
  & = & T 
K_{3}\int_{0}^{\xi}d\kappa\frac{m^{4}}{\kappa^{2}+m^{2}}
\,\,=\,\, 
T K_{3}\left.m^{3}\arctan\frac{k}{m}\right|_{0}^{\xi}
  \label{eq:pth3expanded}
  \end{eqnarray}
where in the second line we have expanded the mass as in (\ref{eq:deltagm}).
Sending the upper integration limit $\xi$ to $\infty$ in 
(\ref{eq:pth3expanded}) changes the result only beyond order $g^3$, 
and we can finally write
\begin{equation}
\left.p_{{\rm 
th}}\right|_{g^{3}}=\frac{m^{3}T}{12\pi}=\frac{T^{4}}{12\pi}g^{3}\,.\end{equation}

\section{Conclusions}

In this paper we have applied the non-perturbative renormalization 
group and its local potential approximation to a 
scalar field theory at finite temperature
to calculate its pressure, screening mass, and scattering amplitude,
covering both weak and strong couplings.
In the perturbative regime, we have shown that the LPA reproduces perturbative
expansions of pressure and screening mass up to and including the order $g^3$.
The latter is important 
as thermal effects at weak coupling are dominated by the effects of 
order $g^3$, and we have indeed verified that the three-dimensional 
flow of the coupling constant is entirely given  by these effects at 
small couplings. The LPA allows a smooth extrapolation of weak 
coupling results into the regime of strong coupling.  Of course, 
because of the presence of the Landau pole, arbitrarily large values 
of the coupling cannot be reached. However the range of values of $g$ 
that can be explored allowed us to demonstrate a clear improvement 
over the strict expansion in terms of the coupling constant. 

We 
have compared the results obtained within the LPA with those of a 
simple 2PI resummation: both methods lead to very similar results in 
the extrapolation to strong coupling. This is not too surprising if 
one considers the diagrammatic content of the two approximations, and 
the fact that both become exact in the large $N$  limit. The LPA has 
the advantage over the 2PI formalism that it yields the one-loop beta 
function correctly (while it is necessary to go to higher order to achieve that in 
the 2PI formalism). This is because the three channels of the 4-point 
function are treated simultaneously in the LPA, albeit approximately 
(loop insertions carry no external momentum in the LPA). 

The NPRG 
provides an understanding of what happens as we move to strong 
coupling. As we have seen two effects compete as $g$ increases: the 
region of three dimensional flow shrinks because the thermal mass 
increases, and the amplitude of the three dimensional flow grows. In 
a sense dimensional reduction continues to play an important role, 
but is no longer related to a weak coupling effective theory. It is 
the competition between these two effects that is responsible, within 
the LPA, for the stability of the results obtained for the pressure 
with increasing couplings: the corrections to the pressure remain 
modest; even for the largest available couplings they never exceed 
10\%.

Of course, the calculations presented in this paper have limitations. 
We have already mentioned the impossibility to go to too large 
couplings in scalar field theory. 
Such a limitation would not appear in an asymptotically free theory like QCD
where many of these difficulties of the perturbative expansion that we have discussed
are encountered (among others). 
Another limitation comes from the 
fact that the LPA ignores the momentum dependence of the self-energy, 
and the effect of the width of quasiparticles is therefore entirely 
neglected. 
Treating such effects is beyond the reach of the derivative expansion.
However the technology to extend the present study in this 
direction exists 
\cite{Blaizot:2005wd,Blaizot:2006vr,Blaizot:2005xy,Blaizot:2006ak}.

\acknowledgments
We would like to thank Aleksi Vuorinen for pointing out
the analytic solution of Eq.~(\ref{eq:integralnoverk}).
Authors R.~M.--G.~and N.~W.~are grateful for the hospitality of  the 
ECT* in Trento where part of this work was carried out.
Feynman graphs have been drawn with the packages
\textsc{Axodraw} \cite{Vermaseren:1994je} and \textsc{Jaxodraw} 
\cite{Binosi:2003yf}.

\noindent{\bf Note:} After this paper was  presented by one of us (A.~I.)
at the third international conference on the Exact 
Renormalisation Group, in Lefkada, Greece, September 18, 2006, 
D.~Litim and J.~Pawlowski informed us about their own investigation on 
this subject \cite{Litim:2006ag}.

\section*{Appendices}
\appendix

\section{Flow equation at finite temperature\label{app:Flowequation}}

\begin{figure}
\begin{center}\includegraphics[%
   scale=0.7]{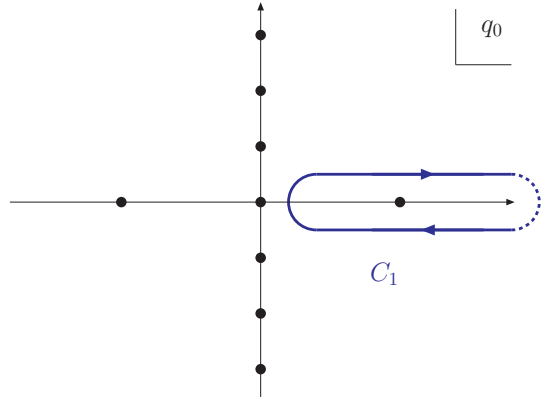}\end{center}
\caption{Integration path for Eq.~(\ref{eq:thermalsum3}). 
\label{fig:integrationpath1}}
\end{figure}

Sum-integrals that occur in various parts of our work can be 
calculated from standard contour integrations using the 
formula:
\begin{equation}
\int\frac{d^{d}q}{(2\pi)^{d}}f(q)
\, \rightarrow \,
T\sum_{n}\int\frac{d^{d-1}{\bf q}}{(2\pi)^{d-1}}f(i\omega_n,{\bf q})
\, = \, 2 \oint_{C_{1}}\frac{dq_{0}}{2\pi 
i}\left[n(q_{0})+\frac{1}{2}\right] \int\frac{d^{d-1}{\bf q}}{(2\pi)^{d-1}} f(q_{0},{\bf q}),
\label{eq:thermalsum3}
\end{equation}
with the Matsubara frequencies $\omega_n=2n\pi T$, 
$n(q_{0})=(\exp(q_{0}/T)-1)^{-1}$
and $n(q_{0})=(\exp(q_{0}/T)-1)^{-1}$ the bosonic distribution 
function. 
Equation (\ref{eq:thermalsum3}) is valid if $f(q_{0},{\bf 
q})$, considered as a function of $q_0$ has singularities only on the 
real axis and is an even function, $f(q_{0},{\bf q})=f(-q_{0},{\bf 
q})$. The contour $C_{1}$ is
displayed in Fig.~\ref{fig:integrationpath1}.
The following useful formulae are easily established:
\begin{eqnarray}
T \sum_n \frac{1}{-(i\omega_n)^2 + \omega_\kappa ^2} & = & \frac{1 + 
2 n(\omega_\kappa)}{2\omega_\kappa} , \label{eq:Tsum1} \\
T \sum_n  \frac{1}{\left(\omega_n^2 + \omega_\kappa ^2  \right)^2} & 
= & \frac{1 + 2 n_\kappa - 2\omega_\kappa 
n'_\kappa}{4\omega_\kappa^3} , \label{eq:Tsum2} \\
T \sum_n  \frac{1}{\left(\omega_n^2 + \omega_\kappa ^2  \right)^3} & 
= & \frac{3}{16\omega_\kappa^5}\left\{1 + 2 n_\kappa - 2\omega_\kappa 
n'_\kappa + \frac{2}{3}\omega_\kappa^2 n_\kappa '' \right\} . 
\label{eq:Tsum3}
\end{eqnarray}

As an application, let us derive Eq.~(\ref{eq:potentialTstart}) for 
the effective potential. Consider the longitudinal modes first.
With the regulator (\ref{eq:LitimIppVacTh}) the ${\bf q}$ integration 
in Eq.~(\ref{eq:basicpotential}) 
can be performed analytically, 
leaving 
\begin{eqnarray}
\partial_{\kappa}V_{\kappa}(\rho) = K_{d-1}\kappa^{d}T\sum_n 
\frac{1}{\omega_n^{2}+V'_\kappa(\rho)+2\rho 
V''_\kappa(\rho)+\kappa^{2}}\,,
\label{eq:potentialT6}
\end{eqnarray}
where $K_d$ is defined after Eq.~(\ref{eq:omegakappaTL}).
When $V'_\kappa(\rho)+2\rho 
V''_\kappa(\rho)+\kappa^{2}>0$, we can use (\ref{eq:Tsum1}) and 
obtain \begin{eqnarray}
\partial_{\kappa}V_{\kappa}(\rho) & = &
K_{d-1}\kappa^{d}\frac{2 
n(\omega_\kappa^L)+1}{2\omega_\kappa^L}\,,\label{eq:potentialT7}
\end{eqnarray}
where $\omega_\kappa^L$ has been defined in (\ref{eq:omegakappaTL}).
The transverse piece that leads to (\ref{eq:potentialTstart}) is 
calculated analogously.

\section{Annotations to perturbative 
calculations\label{app:perturbativeremarks}\label{app:technicaldetails}}

In this Appendix, we collect some technical details of the 
perturbative expansion of the flow equations. We will use the 
following results through order~$g^3$:

The coupling at finite temperature for arbitrary soft $\kappa \leq 
\xi$ follows from (\ref{eq:couplingintegral}) and reads at order 
$g^3$:
\begin{eqnarray}
\left.g_{\kappa}^{2}\right|_{g^{3}} & = & \frac{9 g^3}{\pi^2} \left( 
\arctan\frac{\kappa}{g T} - \frac{\pi}{2} - \frac{g \kappa 
T}{3}\frac{5\kappa^2 + 3 g^2 T^2}{\left(\kappa^2 + g^2 T^2\right)^2} 
+ \frac{8 g T}{3\xi}  + O( (g T/\xi)^2 ) \right) .
\label{eq:gth3gIR}
\end{eqnarray}

The thermal mass has the following behavior for small $\kappa$ which 
can be obtained by
expanding (\ref{eq:mth2g}) and evaluating (\ref{eq:mth3}) for 
arbitrary $\kappa$
\begin{eqnarray}
\left.m_{{\rm th},\kappa}^2\right|_{g^{2}} & = & g^2 \left( T^2 - 
\frac{4 T \kappa}{\pi^2} + \frac{\kappa^2}{2\pi^2} + O(\kappa^5) 
\right),\label{eq:mth2gIR} \\
\left.m_{{\rm th},\kappa}^{2}\right|_{g^{3}} & = & \frac{6 g^3 
T^2}{\pi^2} \left( \arctan\frac{\kappa}{g T} - \frac{\pi}{2} - 
\frac{1}{3}\frac{g \kappa T}{\kappa^2 + g^2 T^2}
+ \frac{4 g T}{3 \xi} + O( (g T / \xi)^2 ) \right) .
\label{eq:mth3gIR}
\end{eqnarray}
The vacuum flow $m_{{\rm vac}, \kappa}^2$ is given in (\ref{eq:diffset3b}).

\subsection{Coupling at finite temperature through order 
$g^4$\label{app:coupling4}}

In order to calculate the constant under the logarithm of the $g^4$ 
coefficient for the
thermal mass, we have to be more careful in integrating the flow 
equation for the coupling
(\ref{eq:diffset1c}). Dividing this equation by $g_\kappa^4$ and 
multiplying by $d \kappa$,
we can integrate the left and right hand sides of 
(\ref{eq:diffset1c}) independently. The r.h.s.~contains IR as well as 
UV problems that we have to cope with.
We introduce an intermediate scale $g T \ll \xi \ll T$ 
(e.g.~$\xi\sim\sqrt{g} \, T$) and calculate this
integral in two pieces.
For $\kappa \geq \xi$, we can neglect $m_\kappa \propto g T \ll \xi$ 
in $\epsilon_\kappa$
and regulate the remaining integral
by subtracting its leading divergent pieces in the IR and UV limits.
Abbreviating $X(\epsilon_\kappa)\equiv 
(1+2n(\epsilon_\kappa)-2\epsilon_\kappa 
n'(\epsilon_\kappa)+\frac{2}{3}\epsilon_\kappa^{2}n''(\epsilon_\kappa))/{\epsilon_\kappa^5}$,
we get
\begin{equation}
  \frac{9}{2\pi^2} \int_0^\infty d\kappa \left[
\kappa^4 X(\kappa)
-\frac{16 T}{3\kappa^2}\theta(T-\kappa) - 
\frac{1}{\kappa}\theta(\kappa-T) \right]
= \frac{9}{2\pi^2}\left(4+\gamma-\log(2\pi)\right)
\label{eq:integralXk}
\end{equation}
with Euler's constant $\gamma = 0.577216$.
This result is obtained by integrating those pieces of $X(\kappa)$ 
which contain derivatives of $n(\kappa)$ by parts and using for 
$n(\kappa)/\kappa$
\begin{eqnarray}
\int_0^\infty \frac{d\kappa}{\kappa} \left( \frac{2}{e^\kappa - 1}
-\frac{2 + \kappa}{\kappa(1+\kappa)} \right) = \gamma - \log(2\pi)
\label{eq:integralnoverk}
\end{eqnarray}
which can be derived by multiplying the terms by $\kappa^a$ and 
analytically continuing the result to $a\rightarrow 0$.
Using (\ref{eq:integralXk}), we can write the $\kappa \geq \xi$ contribution as
\begin{equation}
  \frac{9}{2\pi^2} \int_\xi^\Lambda d\kappa \,
   \kappa^4 X(\epsilon_\kappa)
  = \frac{9}{2\pi^2}
  \left( \frac{16 T}{\xi} + \log\frac{\Lambda}{2 \pi T} + \gamma - \frac{4}{3}
  + O(\xi/T)
  \right)
  + O(g^2),
\label{eq:appcouplingUV1}
\end{equation}
valid for sufficiently small $\xi\ll T$ and sufficiently large $\Lambda\gg T$.

For $\kappa \leq \xi$, we can expand $X(\epsilon_\kappa)$ as in 
(\ref{eq:nbcombinationexpanded}).
In a first step, we use the expression (\ref{eq:couplingintegral}) 
that leads to the $g^3$ contribution of the mass and expand it for 
large $\xi \gg g T$
\begin{equation}
\frac{24 T}{\pi^2}\int_0^\xi d\kappa \frac{\kappa^4}{\left(\kappa^2 + 
m^2 \right)^3} =
  \frac{9 T}{2 \pi m} - \frac{24 T}{\pi^2 \xi} + \frac{T}{m} O((m/\xi)^{3}).
\label{eq:appcouplingIR1}
\end{equation}
It is assuring that the $1/\xi$ terms cancel when adding 
(\ref{eq:appcouplingUV1}) and (\ref{eq:appcouplingIR1}), but we are 
still missing an $O(1)$ contribution which enters on the
one hand through the missing $\kappa$ dependence of $m$ in 
(\ref{eq:appcouplingIR1}),
and on the other hand through the
missing $g^3$ contribution to $m$.
Both can be taken into account by replacing $m^2 \rightarrow 
m_\kappa^2 = m_{{\rm vac}, \kappa}^2 + \left.m_{{\rm th}, 
\kappa}^2\right|_{g^2, g^3}$ using (\ref{eq:diffset3b}), 
(\ref{eq:mth2gIR}), and (\ref{eq:mth3gIR}).
The missing contribution can be thus calculated by
\begin{eqnarray}
\lim_{g\rightarrow 0} \frac{24 T}{\pi^2}\int_0^{\xi\rightarrow 
\infty} d\kappa \left[ \frac{\kappa^4}{\left(\kappa^2 + m_\kappa^2 
\right)^3} - \frac{\kappa^4}{\left(\kappa^2 + g^2 T^2 \right)^3} 
\right] \!\!\!\!\!\!\!\!\!\!\!\!\!\!\!\!\!\!\!\!\!\!\!\!\!\!\!\!
\!\!\!\!\!\!\!\!\!\!\!\!\!\!\!\!\!\!\!\!\!\!\!\!\!\!\!\!
\!\!\!\!\!\!\!\!\!\!\!\!\!\!\!\!\!\!\!\!\!\!\!\!\!\!\!\!
\!\!\!\!\!\!\!\!\!\!\!\!\!\!\!\!\!\!\!\!\!\!\!\!\!\!\!\! & & \nonumber \\
& = &
\frac{432 T^2 }{{\pi }^4} \int_0^\infty d\bar\kappa
\frac{{\bar\kappa }^4}{\left( T^2 + \bar\kappa^2 \right)^5} \left( 
-\frac{{\bar\kappa}^3}{3} +
       T \left( T^2 + {\bar\kappa}^2 \right)
        \left( \frac{\pi}{2} + \frac{\bar\kappa}{T} -
          \arctan(\frac{\bar\kappa}{T}) \right)  \right) + O(g)\nonumber \\
& = & \frac{48}{{\pi }^4} + \frac{27}{8 \pi^2} + O(g),
\label{eq:appcouplingIR2}
\end{eqnarray}
where we have expanded the first denominator around $\kappa^2 + g^2 
T^2$ and substituted $\bar\kappa = \kappa / g$ to form the 
$g\rightarrow 0$ limit.

We obtain the final result by combining (\ref{eq:appcouplingUV1}) + 
(\ref{eq:appcouplingIR1}) + (\ref{eq:appcouplingIR2}):
\begin{eqnarray}
  \frac{9}{2\pi^2} \int_0^\Lambda d\kappa \,
\kappa^4 X(\epsilon_\kappa)
  & = & 
  \frac{9}{2\pi^2}
  \left( \frac{\pi}{g} + \log\frac{\Lambda}{2 \pi T}
  + \gamma - \frac{7}{12} + \frac{32}{3\pi^2} \right)
  + O(g) 
\label{eq:appcoupling}
\end{eqnarray}
Inserting this result into (\ref{eq:diffset1c}), we obtain
\begin{equation}
\frac{1}{g_0^2} - \frac{1}{g_\Lambda^2} = \frac{9}{2 \pi g_\Lambda} + 
\frac{9}{2\pi^2} \left( \log\frac{\Lambda}{2 \pi T}
   + \gamma - \frac{7}{12} + \frac{32}{3\pi^2} \right) + O(g_\Lambda),
\end{equation}
or, expanding and using (\ref{eq:couplingrelationvac}) to change from 
$g_\Lambda$ to $g_\kappa$ for $T\lesssim \kappa \leq \Lambda$, we 
obtain
\begin{equation}
g_0^2 = g_\kappa^2 - \frac{9g_\kappa^3}{2 \pi}
- \frac{9 g_\kappa^4}{2\pi^2} \left( \log\frac{\kappa}{2 \pi T}
   + \gamma - \frac{61}{12} + \frac{32}{3{\pi }^2} \right) + O(g_\kappa^5),
\label{eq:coupling4}
\end{equation}
which connects the value of the coupling at $\kappa = 0$ to the value of the
coupling $\kappa \gtrsim T$.

\subsection{Thermal mass through order $g^4 \ln g$\label{app:mass4}}

When deriving the thermal mass at order $g^3$, in (\ref{eq:mth3}) we 
had neglected
the flow of $g_\kappa$ and $m_\kappa$ at order $g^3$ which may change 
the result
of the mass at order $g^3$. We will see that in the IR region $\kappa \leq \xi$
(with $g T \ll \xi \ll T$)
such corrections would only affect the result at order $g^4 \ln g$, 
thereby justifying
the IR calculation (\ref{eq:mth3}).
In the UV region $\kappa \geq \xi$ on the other hand,
these corrections would affect the result only at order $g^4$,
thereby establishing
(\ref{eq:mth3}) not only as the correct result for the IR, but for 
all $\kappa$ at order $g^3$.

The $g^3$ correction to the coupling only affects $g_\kappa$ in the 
first line of
Eq.~(\ref{eq:mthc}). We can therefore expand $g_\kappa^2 \rightarrow 
g^2 + \delta g_\kappa^2$
with $\delta g_\kappa^2 = g_\kappa^2 |_{g^3}$ from (\ref{eq:gth3gIR}).
(The $g^2$ in the second line of (\ref{eq:mthc}) is not affected by this, as we
subtract a clean $g^2$ contribution.)
In the region $\kappa \leq \xi$, we can expand the thermal 
distribution function in (\ref{eq:mthc}). The correction to 
(\ref{eq:mth3}) is then given by
\begin{equation}
\frac{4T}{\pi^2} \int_0^\xi d\kappa \,  \delta g_\kappa^2 
\frac{\kappa^4}{(\kappa^2 + m^2)^2}
= \frac{96 g^4 T^2}{\pi^4}
\left( \log\frac{g T}{\xi} + \frac{25}{24} + \frac{9\pi^2}{128} - 
\frac{3\pi g T}{4\xi} + O((g T/\xi)^2) \right) .
\end{equation}
This contribution is of the order $g^4 \ln g$. We expect the $\log(g 
T/\xi)$ from the IR region $\kappa \leq \xi$ to combine with a 
corresponding $\log(\xi / T)$ in the UV region $\kappa \geq \xi$, but 
this calculation is beyond the scope of this paper. Since $g^3$ 
corrections to the coupling and mass are IR results stemming from 
$\kappa \leq \xi$, it is clear that no further $g^3$ contribution can 
build up in the UV.
Therefore, there can not be any additional correction to the thermal mass,
and (\ref{eq:mth3}) already captures the full $g^3$ term.

\bibliography{lpa}

\end{document}